# Mawrth Vallis, Mars: a fascinating place for future in situ exploration

François Poulet[1], Christoph Gross[2], Briony Horgan[3], Damien Loizeau[1], Janice L. Bishop[4], John Carter[1], Csilla Orgel[2]

[1]Institut d'Astrophysique Spatiale, CNRS/Université Paris-Sud, 91405 Orsay Cedex, France

[2]Institute of Geological Sciences, Planetary Sciences and Remote Sensing Group, Freie Universität Berlin, Germany

[3]Purdue University, West Lafayette, USA.

[4]SETI Institute/NASA-ARC, Mountain View, CA, USA

Corresponding author: François Poulet, IAS, Bâtiment 121, CNRS/Université Paris-Sud, 91405 Orsay Cedex, France; email: francois.poulet@ias.u-psud.fr

Running title: Mawrth: a fascinating place for exploration

**Abstract**

After the successful landing of the Mars Science Laboratory rover, both NASA and ESA initiated a selection process for potential landing sites for the Mars2020 and ExoMars missions, respectively. Two ellipses located in the Mawrth Vallis region were proposed and evaluated during a series of meetings (3 for Mars2020 mission and 5 for ExoMars). We describe here the regional context of the two proposed ellipses as well as the framework of the objectives of these two missions. Key science targets of the ellipses and their astrobiological interests are reported. This work confirms the **proposed ellipses contain** multiple past Martian wet environments of subaerial, subsurface and/or subaqueous character, in which to probe the past climate of Mars, build a broad picture of possible past habitable environments, evaluate their exobiological potentials and search for biosignatures in well-preserved rocks. A mission scenario covering several key investigations during the nominal mission of each rover is also presented, as well as descriptions of how the site fulfills the science requirements and expectations of *in situ* martian exploration. These serve as a basis for potential future exploration of the Mawrth Vallis region with new missions and describe opportunities for human exploration of Mars in terms of resources and science discoveries.

# 1. Introduction

The major science objectives that drive modern robotic exploration of Mars are to determine whether life arose there, to probe the history and state of the Martian climate, and to prepare for future human exploration. Following the successful investigations by the Mars Exploration Rovers and Mars Science Laboratory (MSL) supplemented by the tremendous data gathered in particular by the Mars Express and Mars Reconnaissance Orbiter spacecrafts, NASA and ESA space agencies have planned to send other rovers to Mars in 2020 in two separate and independent missions. NASA´s Mars2020 mission is based on a MSL-type rover with a distinct instrument suite and an additional catching unit, with the objective of searching for biosignatures and collecting samples toward an eventual sample return mission (Mustard et al., 2013; Williford et al., 2018). The ESA ExoMars rover mission is a Russian-European joint mission, also with the overarching objective to search for biosignatures in surface and subsurface samples (Vago et al., 2017). It was critically important to identify a site that contains compelling science with respect to the mission goals. The space agencies have therefore selected their specific landing site from numerous proposals from different teams based on safety criteria and scientifically promising targets (Grant et al. 2018; Vago et al. 2018).

Among the different landing sites proposed for the two missions, the region of eroded plateaus flanking the Mawrth Vallis channel has been considered as one of the final candidates. Two different ellipses have been proposed on those eroded plateaus to fit the different initial safety requirements given by NASA and by ESA during their respective call for landing sites (Fig. 1). During the various landing site selection meetings, we demonstrated that these ellipses bear the record of multiple ancient aqueous environments as evidenced by the great mineralogical and lithology diversity combined with excellent preservation contexts. Unfortunately, neither ellipse was selected as the final landing site for either mission. In the case of the ExoMars mission, technological constraints hampered the selection of the site, as the proposed landing site at Mawrth Vallis was located roughly 1000 m higher in elevation than the competing Oxia Planum site. Other drawbacks included terrain roughness and slope distribution, although the site was deemed trafficable. This site was also strongly considered for MSL (Michalski et al., 2010), so that we believe that Mawrth Vallis will likely be among top candidates for future missions.

The purpose of this paper is to review the importance and the uniqueness of the Mawrth Vallis region, within the context of current and future in situ exploration of Mars. Using the work performed for the preparation of the past landing site selections, we describe the major characteristics of the site in regards to Mars science and objectives of both forthcoming missions. The site offers the opportunity to investigate a very ancient section of the Martian stratigraphy and to sample rocks from critical epochs of Mars, likely ranging from the Early Noachian up into the early Hesperian. The prevalence of water environments and geochemical gradients that are observed along the strata are key markers to understand Mars's early history and to probe past climates. More importantly, in situ exploration at Mawrth Vallis should provide a critical window into what were potentially the most habitable and well-preserved environments that Mars has experienced.

After a presentation of the regional context, the locations and the major characteristics of the proposed landing site ellipses are described. The major science targets and several astrobiological interests are then discussed. Exploration scenarios compliant with the technical capabilities of Mars2020 and ExoMars-type missions are then proposed to emphasize the accessibility of the terrains and the distribution of the major science targets over a few km². The uniquely diverse astrobiological potential of the Mawrth Vallis terrains are finally compared with future in situ investigations.

## 2. Context

### 2.1. Mineralogical and lithological features/interests

The Mawrth Vallis region is at the boundary of the southern highlands and the northern lowlands near 21-26°N and 336-344°E with elevation varying from approximately −1500 to −3300 m (Loizeau et al., 2007) (Fig.1). This region has received a lot of attention since OMEGA (Observatoire pour la Minéralogie, l'Eau, les Glaces et l'Activité) and CRISM (Compact Reconnaissance Imaging Spectrometers for Mars) discovered extensive (both vertically and horizontally) deposits of diverse and abundant hydrated silicates (Poulet et al., 2005, 2008; Loizeau et al., 2007, 2010; Michalski and Noe Dobrea, 2007; Bishop et al., 2008; Wray et al., 2008; McKeown et al., 2009a; Noe Dobrea et al., 2010) in association with layered terrains first identified with MOC images (Malin and Edgett, 2000). The dominant hydrated phases are Fe/Mg-smectites, Al-smectites, kaolins, hydrated silica, $Fe^{3+}$oxihydroxides and ferrous phases. More recent analyses reported additional localized deposits of sulfates (alunite, jarosite, bassanite), poorly crystalline aluminosilicates (allophane and imogolite), zeolites, mica (celadonite) and possible carbonates (Farrand et al., 2009; Wray et al., 2010; Poulet et al., 2014; Bishop and Rampe, 2016a; Amador et al., 2018). *This region thus contains one of the greatest diversity of aqueous products on the planet discovered so far.*

The alteration mineral assemblage is associated with a thick (>200m) section of layered rocks that are present over almost 2000 m of topographic relief extending over hundreds of km laterally (Loizeau et al., 2007; 2010). These thick and layered deposits also observed in numerous outcrops in the western Arabia Terra region (Noe Debra et al., 2010) are characterized by the same stratigraphy. Two main superposed units are basically present in the deposits: at the top of the section, rocks comprise Al-smectites, kaolins and hydrated silica, while $Fe^{3+}$-smectites dominate underneath (Bishop et al., 2008; Wray et al., 2008; Loizeau et al., 2010). Note however the lower Fe-smectites unit exhibits spectral variability indicating the presence of ferrous phases, mica (celadonite), Fe/Mg-vermiculitic phases sometimes mixed with primary minerals such as pyroxenes for the lowest strata (Poulet et al. 2008; Bishop et al., 2008; Poulet et al., 2014). Of special interest is the presence of exposures of dominantly hydrated phases over most of the preserved Noachian-aged margins of Chryse Planitia including the low stratigraphy of Mawrth Vallis. These minerals have spectral features that are generally similar to ancient altered bedrock elsewhere on Mars, and are consistent with either vermiculites or smectite-bearing mixed-layered clays (Carter et al., 2015a). Local sulfate deposits are also observed, with jarosite, possible copiapite, and alunite present at the top of the sequence (Farrand et al., 2009; 2014, Bishop et al., 2018a), and bassanite identified in lower layers (Wray et al., 2010). Finally, carbonate detection associated with the light-toned outcrops have been recently reported (Amador et al., 2018; Bultel et al., 2019). The different secondary phases found in the region are listed in Table 1, while their vertical distribution associated to schematic representations of layers and fractures is shown on Fig. 2.

The hydrated layered light-toned outcrops are capped by a regionally-extensive dark mesa-forming unit that exhibits unaltered mineral spectral signatures (Loizeau et al., 2007). From crater count, the cap rocks are 3.7-3.6 Gy old (Loizeau et al., 2012), which indicates that surface aqueous alteration stopped prior to the age of this dark and non-altered materials. Indeed, the phyllosilicate-bearing rocks may be even more ancient. Crater counting suggests that the oldest light-toned, smectite-bearing rocks in this region are Early- to Mid-Noachian (Michalski and Noe Dobrea, 2007, Michalski et al., 2010; Loizeau et al., 2012).

The bedrock in the Mawth Vallis region is finely layered with strata thicknesses of 1 m or less indicating sedimentary processes (Loizeau et al., 2007; 2015). The lack of diagenetic maturity in these layered sediments implies that the original chemical, mineralogical, and textural properties of the rocks are likely to reflect ancient conditions during the time period in which they formed (Michalski et al., 2010). T*he Mawrth Vallis region thus enables investigations of some of the most ancient outcrops of sedimentary, altered rocks on Mars over varying alteration episodes, spreading over a few 100s millions years between the early Noachian (Phyllosian) and the global transition into the Hesperian (Theikian).*

Evidence of past fluvial activity (ancient valleys and ponds) is visible on many outcrops of the region's plateau (Fig. 3). Numerous km-long to tens-of-km-long fluvial valleys incise the surface of the clay-bearing unit, generally indicating ancient flows towards Mawrth Vallis (or a former valley preceding incision of the outflow channel) and towards craters, like Oyama crater, around the ellipses (Fig. 3). The valleys are all more or less filled by the dark capping unit. A fluvial branch forming an amphitheater-shaped valley head can be observed in the eastern part of the ExoMars landing site (Fig. 3D). This feature could indicate a formation by groundwater sapping and may also indicate ponding in topographic lows and regions with poor drainage. This zone is covered by capping unit material so that the floor may not be accessible, but the relationships of the fluvial activities with the formation of altered rocks will be a fundamental aspect to be studied in this region.

Other fluvial features are strongly eroded and often appear as inverted relief (Loizeau et al., 2007; Mangold et al., 2010; Loizeau et al., 2015). These geological features actually support several aqueous processes from early fluvial activity possibly during the layers' deposition period to a later formation after the Mawrth Vallis channel. *The region offers the advantage to determine the duration and timing of several surface water episodes responsible for the formation of these liquid water-related erosional and depositional features.*

Additional intriguing geological observations in the Mawrth Vallis region are various linear structures found in some portions of the Fe/Mg-unit (Michalski and Noe Dobrea, 2007; Wray et al., 2008; Loizeau et al., 2012; Loizeau et al., 2015). These structures are typically a few tens of meters large and from 100 m to several km long (Loizeau et al., 2012). They are observed as fractures infilled by dark material, or ancient inverted fractures. Some of those features are surrounded by a brighter halo, sometimes forming parallel ridges (Fig. 4B), comparable to the halo-bounded fractures observed by Okubo and McEwen (2007) in Candor Chasma and the ridges observed in the Nili Fossae area (Saper and Mustard, 2013; Pascuzzo et al., 2019). Others have dark centers probably due to sand filling, and they are several tens of meters to 100s m long (Fig. 4B). The densest set of these dark fractures is observed on the flank of Mawrth Vallis in deeper-seated layers (Loizeau et al., 2015). Similar dense fracturing is also observed in the vermiculitic-unit of the Oxia Planum landing site for ExoMars (Quantin et al. 2018b), which could suggest a common origin between these layers.

Other fractures in Mawrth Vallis are bright white, often with dark centers (Fig. 4C and 4D). These fractures tend to appear in sharp contrast to surrounding surface sediments, and sometimes terminate local bedforms, both suggesting that they are raised above the surrounding terrain. These bright fractures tend to occur closer to the top of the sequence (Loizeau et al., 2015), and their morphology may be correlated with depth. In many locations, bright networked rectangular fractures are observed eroding out from immediately under the capping unit (Fig 4D), while more exposed areas in lower stratigraphic sequences exhibit more isolated long curvilinear and often branching bright fracture morphologies (Fig 4C). The raised and bright nature of these fractures suggests mineral precipitation in the form of e.g., sulfates or silica, although these fractures have yet to be correlated with CRISM mineral detections. The different morphologies could be related to conditions during fracturing (Kinzelman & Horgan, 2019). Rectangular fractures typically imply contraction at the surface, often by wetting/drying cycles, followed by sediment infill or mineral precipitation in the joints (e.g., Kocurek & Hunter, 1986). A surface origin of the rectangular fractures at Mawrth could be consistent with their location at the top of the stratigraphy. In contrast, both the filled and unfilled fractures at Mawrth lower in the stratigraphy often exhibit curvilinear and branching morphologies. This type of fracturing may instead suggest subsurface contraction without drying, as can occur during burial and dewatering of saturated clay-rich sediments (e.g., McMahon et al., 2016; Siebach et al., 2014), perhaps consistent with subsurface dewatering of the clays under saturated subaqueous or saturated aquifer conditions.

Finally the top of the clay sequence also shows fracturing at different scale. This fracturing is characterized by regular, polygonal blocks ~1 m large (mostly in the Al-clay rich part of the sequence) creating an elephant-skin texture (north part of Fig. 4A) or as irregular fractures, creating blocks from <1 m to ~10 m in width (mostly in the Fe-clay rich part of the sequence).

The mapping of these >10s meter-long fractures and ridges is a complex and very time consuming task, so that the mapping of specific types of fractures was systematic to the Mars2020 ellipse only, and roughly performed to some other particular places in the much larger ExoMars envelope (Kinzelman & Horgan, 2019).. They are an outstanding question in terms of formation processes as running hypotheses for their formation include crater-related faults, desiccation or hydraulic fracturing for their initial formation, sometimes followed by underground fluid flow, including the possibility for low-temperature hydrothermal process. Hydraulic fracturing during the burying of the sediments has been proposed for Yellownife bay and Murray formation in Gale (Caswell and Milliken, 2017). A similar process could explain the fractured basal unit present in the lower part of the stratigraphy. The halo-bounded fractures, ridges and inverted fractures support fluids circulation leading to the precipitation of minerals and indurated material inside fractures and in the rocks close to the fractures. However, we hypothesize that the rectangular networks of smaller filled fractures at the top of the sequence likely formed due to desiccation of wet sediments in a sub-aerial environment (similar to mudcracks) or of the swelling-clays present in the rock. The smallest scale "elephant skin" fracture texture in the Al-clay unit may either be due to desiccation or to thermal contraction at the surface. *The diverse fractures morphologies at Mawrth Vallis suggest that the site has preserved multiple subsurface and surface environments.*

## 2.2. Alteration environments

*The extremely diverse mineralogy and lithology described in the previous section suggest multiple aqueous deposition and geochemical environments with subsurface and surface water settings starting prior to ~4.0 Ga and ending 3.7–3.6 Ga ago.* These environments have been discussed in several papers including for instance Michalski and Noe Dobrea (2007), Bishop et al. (2008), Michalski et al. (2010), Loizeau et al. (2012), Bishop et al. (2013), Bishop and Rampe (2016a), Bishop et al. (2018a, 2018b). These studies were mainly based on the fact that mineralogical compositions could serve as fingerprints to constrain the past environmental conditions under which they formed. We below review these environments.

- The abundant thick (~100-200 m) Fe/Mg-smectite deposits were likely formed via aqueous alteration of mafic to ultramafic rocks or subaerial deposits requiring moderate water/rock ratio (W/R hereafter) environment within a closed-system (Baker, 2017), and show no evidence of thermal processing such as metamorphism. The layering of the Fe/Mg smectites unit is complex indicating perhaps multiple depositional events over a prolonged and continuous period of time. Several water settings can be envisioned: marine/palustrine environments and/or pedogenesis in a moist climate and/or aqueous transport of sediments rich in detrital clay minerals. For instance, the presence of celadonite suggested in some strata of the deposits could indicate submarine weathering of volcanics, or hydrothermal fluid with temperatures between 20°C and 60°C (Poulet et al., 2014). The celadonite/nontronite assemblage is however also found in pedogenic settings, resulting from a simple transformation of celadonite to smectites during soil formation (Reid et al., 1988; Righi and Meunier, 1995). Surface activity is also supported by the remnants of ancient channels. On the other hand, if the alteration occurred due to some localized hydrothermal phenomenon (e.g., associated with an igneous intrusion), we would expect to see alteration products tied **only** to fluid conduits, such as faults and fractures, and we would not expect to see such a widespread and consistent distribution of alteration minerals through the large Mawrth Vallis flat-lying homogeneous units. Other such hydrothermal features have been identified from orbit elsewhere on Mars, especially in the Nili Fossae area (Pascuzzo et al., 2019). A strictly hydrothermal interpretation for the origin of the mass of altered, layered sediments is therefore very unlikely.

- Changing redox conditions occurred at the surface or subsurface to explain the observed $Fe^{2+}$-rich unit at the boundary of the Fe/Mg-smectites and the upper Al/Si-rich minerals: leaching and/or organic matter and/or salinity can be a reducing agent for smectites (Bishop et al., 2013).

- Mildly acidic conditions and/or warm and humid climates are hypothesized at the top of the section as those environments support active hydrolysis and ion leaching that can result in formation of kaolinite and hydrated silica (Bishop et al., 2013). Nanophase-aluminosilicates are commonly detected in well-drained mildly acidic soils derived from volcanic ash containing feldspar and pyroxene (Bishop and Rampe, 2016a), and further weathering of these phases produces halloysite or kaolinite (Ziegler et al., 2003; Rasmussen et al., 2010).

- Evidence for salty environments is present in evaporitic residues found in thin horizons between and above the thick, lower Fe/Mg-smectite unit and the upper Al-phyllosilicate unit. These sulfates likely formed in surface evaporitic environments that were later buried in some cases. Some of these sulfates are Fe/Al-sulfates, suggesting oxidation of sulfides to form locally acidic environments with strong redox gradients (Michalski et al., 2010; Farrand et al., 2014).

- The sequence consisting of Al-rich clay minerals over Fe/Mg-rich clay minerals shows that weathering of rocks from top to bottom was the most likely process at play at the top of the sequence, as indicated by an evolution of the pH from neutral/alkaline to acidic, and closed-system to open-system alteration through the upper stratigraphy of Mawrth Vallis outcrops (Carter et al. 2015b). The mineralogy of this weathering sequence is directly related to climatic and environmental conditions. Typical neutral, well-drained environments are dominated by phyllosilicates produced through hydrolysis by carbonic acid, but the additional presence of sulfates, oxides, reduced iron, and carbonates can indicate other conditions, including variations in water saturation, redox state, and pH. The presence of carbonates recently reported in the well-preserved weathering profiles on Mars including at Mawrth Vallis is consistent with climate models suggesting a denser $CO_2$-rich atmosphere (Zolotov and Mironenko, 2016; Bultel et al., 2019). Finally, the widespread distribution of these profiles on Mars suggests a planetary scale weathering that affects the surface of Mars at least at mid-latitude (Carter et al., 2015b). This clay-rich material could have formed at the surface during short-lived climate excursions (each tens of thousands of years) spread over hundreds of millions of years (Bishop et al., 2018b).

In summary, all the observations point to the formation of Fe/Mg-rich phyllosilicates, a transition to acidic alteration phases and sulfates during periods of wet/dry cycling, crystallization of Al-rich phyllosilicates through pedogenesis or leaching under $CO_2$-rich atmosphere, and persistence of np-aluminosilicates and hydrated silica marking the end of wet conditions on Mars. Layered outcrops observed at Mawrth Vallis are thicker than in other regions of Mars, and may represent processes that took place on a wider scale in wet areas of the planet during the Noachian, but which are better preserved and exposed at Mawrth Vallis.

As discussed in the previous section, the formation of the diverse fractures suggests several possible formation processes including desiccation in a sub-aerial environment, fracturing during burial and dewatering of subaqueous sediments, sometimes followed by subsequent mineral precipitation. This may be explained by the presence of an ancient aquifer circulating at this level in the ancient subsurface. Fluid circulation was still happening, at least in the subsurface, after the formation of the fractures, thus well after the cementation of the layers. From remote sensing observations, we cannot exclude low-T hydrothermal systems related to this fluid circulation. Although most of these veins are found in the clay-rich deposits away from visible craters, some are found related to crater rims of a few km-size craters and could also correspond to possible impact-induced hydrothermal systems.

### 2.3. Depositional history

*The hydrated sediments at Mawrth Vallis are thus unique on Mars because their formation involves numerous Noachian aqueous environments that sample various alteration pathways and depositional processes.* The history of deposition has been described in Loizeau et al. (2012). Here we refine this history based on the identifications of new minerals and recent works on the region. The buildup of the thick stratigraphy as seen today can be divided into 6 phases:

A. Progressive deposition and alteration of clay sediments (Fig. 5A). The bottom part of the stratigraphy was emplaced from early to middle Noachian. The mineralogy assemblage is dominated by $Fe^{3+}$/Mg smectites, possibly sulfates in some strata and $Fe^{3+}$ oxy-hydroxides. Anhydrous minerals such as plagioclase and pyroxenes are also mixed in some lower basal units (Poulet et al., 2008). As discussed previously, the mineralogy is consistent with low temperature alteration, sustained aqueous activity and moderate W/R. The possible presence of sulfates in the stratigraphy may suggest drier cycles, as commonly observed in thick terrestrial sedimentary records. The material forming the layered rocks could have been altered before being transported and deposited in the region of Mawrth Vallis, through aqueous transport and sedimentation, or the layered rocks/ashes could have been altered in situ, during or after the deposition, in a marine environment or through pedogenesis in a moist climate. In this authigenic hypothesis, the original material could be of fluvial and/or aeolian origin, or pyroclastic. Hence, both surface and subsurface water settings could have been existed: marine environments and/or pedogenesis in moist climate and/or aqueous transport of sediments (Michalski et al., 2010; Loizeau et al., 2012).

B. Progressive deposition and alteration of clay sediments and hydraulic fracturing due to burial (Fig. 5B). These deposits exhibit a large content of hydrated phyllosilicates, which suggests that the rocks are mature sedimentary rocks (Poulet et al., 2008; 2014). The presence of celadonite could be in favor of a marine environment. While progressive deposition continues as in Fig. 5A, the progressive burial of the lower part of the stratigraphy could have led to the formation of fractures in the basal part as the result of hydraulic fracturing. It has been proposed that this lower stratum where the densest network of fractures is visible could correspond to the same deposits as in Oxia Planum (Quantin-Nataf et al., 2018b). Possible circulation of fluids at this stage in the lower part of the stratigraphy could have led to additional alteration in the deeper clay unit.

C. Deposition of the top section, fluid circulation in fractures, fluvial activities (Fig. 3C). The top section was deposited/modified during late Noachian/early Hesperian. Various local geochemical environments are indicated by the diverse mineralogy: kaolins, ferrous clays, poorly crystalline Si-phases, npFe-Ox, Al/Fe$^{3+}$ smectites, carbonates. The water setting was likely surface or subaerial with multiple environments revealing a wet climate with $CO_2$-rich atmosphere: pedogenesis, palustrine, fluvial and ponding settings. Meanwhile, mineralization of the fractures occurred through underground fluid circulation, possibly under low-T hydrothermal activity.

D. Deposition of sulfates/aluminosilicates during dry periods (Fig. 5D). Salty outcrops were deposited throughout the light-toned, layered phyllosilicate unit as tiny pockets of sulfates intermixed with Al-phyllosilicates, or as evaporites on top of the clay sequence formed in ponds or at springs. The timing of these acidic depositions could thus have been multiple, and not necessarily at the end of the pedogenesis event that created the top section. Then, a short-lived, colder climate is the best interpretation to explain the altered rock in the form of np-aluminosilicates without progressive alteration to crystalline clay minerals.

E. Deposition of the capping unit (Fig. 5E). The dark-toned, pyroxene and plagioclase-bearing capping unit was finally emplaced at ~3.6 to 3.7 Gy on the top of the phyllosilicate-bearing units. This capping unit is likely a pyroclastic and windblown deposit, given its draping morphology, its relatively weak thermal inertia compared to other geologic units of the region, the observation of pyroxene on its surface, and its dark color (Noe Dobrea et al., 2010; Loizeau et al., 2015). No more surface aqueous alteration occurred after this deposition.

F. Wind erosion. The whole section is progressively and continuously exhumed during the Hesperian and Amazonian. The erosion was not uniform. Locally, the capping unit has eroded into talus and loose surface material that probably have fed local sand dunes. But in other locations, the phyllosilicate package is totally exhumed or only visible through small windows inside the capping unit. Overall the site was very well preserved by this capping unit hence providing the best-preserved clay-bearing stratigraphy of the northern hemisphere of Mars.

3. **Ellipse properties**

In this section, we propose to document and archive the major characteristics of the ellipses chosen for Mars2020 and ExoMars missions. The overarching goals, the technical capabilities and the scenarios of these two missions are significantly different, which led to the selection of two ellipses presenting different characteristics.

## 3.1. Overview

The proposed landing ellipse for Mars2020 is situated directly on a large exposure of phyllosilicates (Fig. 6) that occurs within layered, fractured bedrock. This area was chosen as a combination of safety criteria and scientific interests as it is one of the largest exposures of the bright, clay-rich bedrock, covered by very few remnants of the capping unit. This maximized the chance to land directly on the most desired geological unit for exobiological goals. It is very close to the ellipse that had been proposed for MSL (Loizeau et al., 2015) but to avoid landing hazards linked to a crater, it has been suggested by the Mars2020 project to shift the ellipse towards the east by a few km. The landing ellipse of 12 km x 10 km contains the two main sub units of the clay sequence, and the transition between those sub-units is accessible in many areas throughout the landing ellipse (Sections 3.2 and 3.3).

Several ellipses for ExoMars with different centers covering the 2020 opportunity launch and azimuth ranges of Mawrth Vallis were studied. They were chosen to fulfill the safety criteria of the EDL (Entry, Descent, Landing) system. Due to the large size of the ellipses, each measuring 120 km by 19 km, the only possible safe region was found south of the Mawrth Vallis channel located on the plateau of a thick stratified accumulation of hydrated minerals (Fig. 7). Ellipse centers were selected to optimize access to a large variety of hydrated materials. The sheer size of the landing ellipses yields a large variety in landscape and landforms such as plains, valleys, mesas, craters and dune- or sand patches and with this a high diversity in accessible outcrops materials.

### 3.2. Geological features

#### 3.2.1. Mars2020 landing ellipse

The surface of the landing ellipse defined for the Mars2020 mission is mostly covered by outcrops of various hydrated minerals (simplified as clay-rich unit in Fig. 8). The lithology of this unit exhibits the typical light-toned, fractured, finely layered rocks (Fig. 9A-D). The clay-rich outcrops are partly covered with fine sediments, light toned to dark toned dust and sand, which form dark aeolian ridges in places (e.g. Fig. 9B and 9F). Remnants of the dark capping unit are also visible. Erosion of the capping unit likely produces dark sand that creates dark talus around the capping unit remnants (Fig. 9E) and probably feeds the dark aeolian ridges. Dark sand also fills any craters, crevasses or holes, making fractures usually appearing dark.

Within the clay-rich unit, erosion reveals the presence of some concentric layered features interpreted as inter-layered craters (Fig. 9F). Those ancient, "ghost" craters could have been created during the layer deposition, hence were filled with layered rocks, and appear today as flat or convex circular layered features, similar to the "ghost" craters of the paleo-surface described in Loizeau et al. (2010), but less densely concentrated than for this paleo-surface. Erosion has also revealed an eroded inverted channel (Loizeau et al., 2015), under the form of a curvilinear series of round to elongated remnants buttes that separates in two branches and merges again.

Several fractures are observed in the clay-rich unit. First of all, most of the clay-rich outcrops are fractured in blocks large from ~5 m to less than 50 cm. This small scale fracturing appears sometimes regular, creating polygonal terrains (Fig. 9C) associated to Al/Si-OH phases. It is more irregular in size and shape most of the time and especially when found in the Fe/Mg-smectite unit (Fig. 9D). Longer fractures are also present, they can reach several hundred meters in length. They are thin, sometime not resolved with the HiRISE images, but are visible because they are darker than the surrounded rocks, probably due to dark sand filling (Fig. 9G).

Larger fractures are also present. Some are present in dense networks (Fig. 9H), and others are mapped as "ridges and halo-bounded fractures". This last category groups more isolated ancient fractures, but are more modified. Among these ridges and halo-bounded fractures are present dark ridges, which are well different from modern aeolian ridges (Fig. 9I). Those ridges can reach >2 m in relief, for widths <5m, notably in the north-western part of the ellipse. Other fractures of this category are large and show modified contours. The central, dark part of those fractures can reach >10m in the south-eastern part of the ellipse (Fig. 9J), but there is a large population of short (50 to 100 m long), large (2 to 4 m large) dark lines in the middle of the ellipse, some presenting ridges, some presenting brighter halos around them (Fig. 9K). This last category of ridges and halo-bounded fractures has a strong interest in terms of aqueous history and habitability of the site as they could involve fluid circulation in fractures and mineral precipitation in the fractures and their vicinity. As suggested before, they might be the sign of a past active hydrothermal system that triggered this fluid circulation.

### 3.2.2. ExoMars landing ellipse

Within the ExoMars landing envelope a large variety of surface features can be found. Due to its larger size than the Mars2020 ellipse, most of these features described in the Mars2020 subsections are obviously present. The main geological units are represented by the clay-bearing light-toned deposits and by the capping unit (Fig. 10). The latter displays a mainly continuous layer in the northwestern part of the landing envelope, whereas a more dissected appearance is observed in central parts. Towards the south, the capping unit appears to thin out and more eroded surfaces can be recognized. Such texture has been mapped as a geological unit named eroded capping unit. In general, the visual appearance of the light toned clay-bearing units present at the ExoMars landing site is comparable to the description in section 3.2.1. As noted by Wray et al. (2009) and Bishop et al. (2008), the morphology at meter scale reveals different sub-units well consistent with the compositional units (Fig. 11): the Fe/Mg-smectite unit usually exhibits a dense fractured polygonal surface, while the units containing Al-clays/hydrated silica is characterized by a smoother texture with sometimes the presence of fracture networks similar to an elephant skin. Each unit also has a distinct color in HiRISE/HRSC images, as observed elsewhere in the Mawrth Vallis region (Loizeau et al., 2010; Bishop et al., 2008). In general, units with Al-clays appear bluer than reddish units with Fe/Mg-phyllosilicates. In many places, layering can be identified within the light-toned clay-bearing unit. For example at the flanks of impact craters, in smaller depressions or in flat regions with little dust/sand coverage like within the southeastern plain shown in Fig. 11.

In the capping unit, some smaller impact craters have exhumed underlying light-toned deposits making an easy access to the clay-rich materials. Also, erosional windows can be found everywhere on the capping unit (Fig. 12), revealing underlying materials accessible by a rover.

From north to south, numerous valleys crosscut the studied envelope (Fig. 3). The largest one in the central area, partly revealing layered, light-toned deposits at the flanks. Some parts of those valleys are inverted: when the terrain is more eroded (southern part of the ExoMars ellipse, for example directly south of spot number (4) in Fig. 10, or at the bottom of the Mawrth Vallis outflow channel), then the valley appears as an elevated feature, in the continuity of the valley depression. These valleys and inverted valleys all show an infill by dark capping unit material. As reported in section 2.1, we also observe a large ponding geological feature in the eastern part of the landing site. This zone is covered by capping unit material so that the floor may not be accessible, but the timing of the formation of altered rocks with the fluvial activity will be a fundamental aspect to be studied for estimating the timing of aqueous activity and duration of the habitability in this region, and hence on Mars.

Finally, the clay-rich layers in the envelope host the diverse suite of large fractures and fracture networks as for the Mars2020 ellipse (Fig. 4 & 9). The fractures do not all exhibit evidence for fluid interaction. Perhaps the most interesting fractures are those that exhibit very bright margins or halos (Fig. 4C-D). These exhibit diverse morphologies: linear/ irregular/curvilinear, sharp or more diffuse boundaries, clear central fracture, sometimes raised above the surroundings suggesting resistant mineral infill.

### 3.3. Mineralogy

#### 3.3.1. Mars2020 landing ellipse

The clay-rich unit at Mawrth Vallis is not uniform in composition. OMEGA and CRISM-targeted observations clearly distinguish the two main sub-units previously mentioned in section 2.1 at the regional scale. CRISM-targeted observations of the landing ellipse reveal a remarkable diversity of minerals (Fig. 13 & 14, Table 1), which is well representative of the top ~50 m sequence of the stratigraphy. All the common phases are present: $Fe^{3+}$/Mg smectites, $Fe^{2+}$/Mg-clay signature, Al-OH, Si-OH, sulfate. As shown in Poulet et al. (2014), the Fe/Mg-rich phyllosilicate subunit contains a large abundance of nontronite (30–70%) mixed with a ferrous mica (<20%). The presence of ferrihydrite is also required to adjust the intensity/shape of the 1.9 µm and 2.3 spectral features. Spectral modeling of the Al-rich phyllosilicate unit reveals a mineral mixture more complex than for the Fe/Mg clay unit with the presence of several hydrated phases: kaolinite, Al/Si-OH phases, and possibly opal.

A specific signature identified as a doublet-type band near 2.22-2.23 and 2.26-2.27 µm is visible in the ellipse. These additional features have been already reported in other locations (Bishop et al., 2013; 2016b; 2018a). They are roughly similar to the positions of the jarosite bands, although the relative intensity is inconsistent with jarosite, and other diagnostic jarosite features are missing. Spectra of acid-treated Fe-smectites exhibit a similar doublet and thus provide a possible explanation for this observed doublet features in some of the upper Al/Si-type unit. These different minerals detected in the NIR seem to follow a vertical sequence at the first order, but horizontal heterogeneity is also observed as some minerals are not detected systematically when erosion exposes the clay sequence. For example, alunite, at the top of the sequence, or kaolinite, silica and the doublet-phase seem to be present in patches.

The particular interest of the Mars2020 proposed landing zone was the access to this large mineral diversity throughout the landing ellipse: wherever a rover would land within this landing ellipse, an access to the whole identified clay sequence would be available within just a few km without the necessity to cross the whole landing ellipse.

### 3.3.2. ExoMars landing ellipse

Less than 10 targeted CRISM observations exist within or at the border of the large ExoMars landing envelope. Additionally, OMEGA and CRISM MSP and HSP data were used to investigate the mineralogy of the region (Fig. 15). The specific minerals present within the ExoMars ellipse are listed in Table 1 and a description is given in section 2.1. In such cases, mineralogical identification at local scale is performed using HiRISE/HRSC color image and/or textural properties of surface and analogy to other deposits in the Mawrth Vallis area. The deposits present within the landing envelope cover the whole upper part of the stratigraphic stack of the Mawrth Vallis deposits shown in Fig. 2. Starting with the capping unit, through the Al-rich deposits, the transition zone with kaolins is omnipresent, as are the Fe/Mg-rich deposits which appear in higher frequency towards the south. Throughout the site, smaller amounts of sulfate-rich material including $KFe^{3+}$-jarosite, Ca-bassanite and KAl-alunite are also detected (Carter, personal communication). The sulfates are always intermixed with the leached top sequence of the stratigraphy. There are rare occurrences of sulfates within the ellipse. These minerals have been reported previously or concurrently within the larger Mawrth Vallis plateau deposits by Wray et al. (2009), Farrand et al. (2009) and Sessa et al. (2018). These are hence not new detections but confirm that the units within the ellipse are of the same mineralogical makeup as the rest of the plateaus.

Just outside and north of the envelope when going down to the scarp of the Mawrth Vallis channel, a paleosurface possibly inter-layered with the nontronite exhibits the presence of $(Al, Fe^{3+}, Fe^{2+}, Mg)$ vermiculitic smectite. This mineralogy is very similar to the one observed in the Oxia envelope (Quantin-Nataf et al., 2018a; 2018b). This supports an additional evidence for a genetic link between the two sites with Oxia-type material present in the lower portion of the Mawrth Vallis stratigraphy. This considerable compositional variety found within close proximity demonstrates again the uniqueness of the Mawrth Vallis region. Despite the large extent of the landing site with an EDL system such as the ExoMars one, a rover could have reached prime mineralogical targets within a 1 km drive as demonstrated in sections 3.4.2 and 5.2.

### 3.4. Science targets

#### 3.4.1. Mars2020 landing ellipse

The main scientific goals of the Mars2020 mission are linked to the return of a set of samples from the site surface and rocks. Science targets are thus selected with respect to the goals of the Mars2020 mission but they are generic of the site. Targets to collect those samples include astrobiological relevant ancient environments, rocks to characterize the geologic record, and samples that may yield to other significant scientific discoveries. In this respect, the proposed landing ellipse at Mawrth Vallis **gives** access to a variety of targets as schematically mapped in Fig. 16 and described below:

1. The deeper Fe/Mg-clays outcrops, of fractured, finely layered bright rocks. Those rocks give access to an ancient environment with moderate water/rock ratio, and a sustained aqueous activity. Those environments may have corresponded to low temperature marine margins and/or surface pedogenesis.

2. The upper Al-clays/Si-OH outcrops, of polygonised brighter rocks. Those rocks are consistent with a higher water/rock ratio and possible acid leaching of the surface and near-surface. Local mineralogy points to well-drained rocks at the top of the sequence and local precipitation of evaporates.

3. The clay sequence transition offers a diversity of ancient habitable environments with a spatial gradient of pH and redox conditions, and in particular a continuous record of wet environments in the middle to late Noachian with potential access to the transition to the Hesperian;

4. Rock locations **(ferrous iron layer)** that recorded particularly reduced environments that offer preferred targets for past habitability and preservation of potential biosignatures;

5. Ridges and halo-bounded fractures that gives potential access to late aqueous activity and hence a different phase in the site habitability;

6. A possible inverted channel at the top of the sequence that offers an ancient surface wet environment with erosion and deposition of transported material;

7. Igneous capping-unit remnants that provide a point in the igneous history of the planet and a datable regional surface.

Most of those science targets are each well distributed within the landing ellipse; a few km-long mission would thus give access to all those targets.

### 3.4.2. ExoMars landing ellipse

The primary objective of the ExoMars mission is to search for signs of past or present life on Mars and to characterize the water/geochemical environment as a function of depth in the shallow subsurface (Vago et al., 2017). Due to the very large size of the landing envelope, only a set of locations is here presented, depicting the various environments that can be found in the different regions of the envelope (Fig. 17). These locations have been selected to be representative of the morphologies observed at HRSC scale. For the success of a mission such as ExoMars-type it is crucial to reach scientifically valuable targets in close proximity to the touchdown point in order to maximize time spent with scientific investigation rather than by driving towards potential scientifically interesting targets. The selected areas possess the major science targets of the region (Table 2). Even if the center of the zone is not on clay-bearing light toned deposits (example 3 for instance), a short drive allows to reach them thanks to the availability of outcrops. Despite different surface characteristics at the landing site, the required science objectives can be thus realized within short roving distances and within the nominal mission time. To emphasize this accessibility, we colour in the landing envelope the parts for which a given touchdown point is never more than 3 km away from a prime target (Fig. 18). The characterization of the accessibility was performed using a CTX mosaic with ~7 m/pixel resolution. The defined prime targets are: Al- and Fe-/Mg- rich deposits (layered materials), fresh light-toned material exposed directly under thin capping unit often exposed by small impact craters and erosional windows, all types of fractures and ridges, fluvial features, such as valleys and ponding features, sulfate- and possible carbonate deposits. At close inspection, only two small regions in the eastern and in the western part of the ellipse are outside a 3 km range. Otherwise cap rock material does show impact craters, or erosional windows that could provide a glimpse into the subsurface.

A quantification has been provided to the ExoMars project (Poulet et al., 2017). The probabilities to reach major science targets (clay outcrops, veins and fluvial features) assuming a drive of 3 km are found to be 1, >0.4 and ~0.5 respectively. As in situ investigations increase the likelihood to detect interesting targets (as demonstrated by the MSL mission), it means that the probability values correspond to lower limit. The probability for the mineralized veins are likely underestimated because 1) these features are identified with HiRISE color strips only that provide a limited partial spatial coverage of the region and 2) their mapping is also not systematic. As these features are associated to the clay-bearing units, the probability should be near the probabilities of reaching clays.

Several hazards were reported by the ExoMars project (mostly TARs, slopes) and could impact the derived probability and the accessibility of the high priority targets. The valleys are very large structures so that we consider the hazards shall have a minor impact on the derived probabilities. However, we cannot exclude that some parts (especially rims) of these features could be hard to reach. The mineralized veins/fractures were roughly mapped with HiRISE images only. As the HiRISE coverage of the region is partial only, the number of fractures and therefore the derived probability values are thus clearly underestimated. We actually estimate that the real probability should be similar to the values derived for clay-bearing outcrops as the veins are associated with these outcrops. There are in general easily accessible, while sand dunes could be problematic to reach some of them and/or their center.

4. **Astrobiological interests of science targets**

From knowledge of terrestrial environments, a site or a science target can be considered as an astrobiologically relevant environment if this environment is prone to respect three broad conditions concerning: 1) habitability, 2) presence of chemical biosignatures and 3) preservation. The highest astrobiological interest occurs at the intersection of these factors, which need to be sought on Mars (Fig. 19). Orbital observations establish geological context, identify and map the different rock units that may represent a diversity of paleoenvironments. In this regard, they are essential to constrain the habitability and preservation criteria. On the other hand, assessing potential presence of chemical signatures at a site requires a rover that can navigate the terrain to conduct complex chemical, isotopic, mineralogical and morphological analyses of a range of targets at multiple spatial scales. Before in situ observations are available, this criterion can be roughly evaluated by using terrestrial analogues. Below, we discuss the suitability of the main science targets of the Mawrth Vallis in terms of habitability, preservation and biomarker potential, sometimes through terrestrial analogy (Table 3).

### 4.1. Habitability potential

As detailed in Hoehler (2007), Mustard et al. (2013) and Hays et al. (2017), the assessment of habitability consists in examining the geologic record so as to identify several factors required as necessary for being a habitable environment: availability of raw materials (CHNOPS elements and a source of electron donors), availability of free energy in sufficient abundance and adequate form, availability of liquid water (a solvent, catalyst, and source of energy in some environments) and favorable conditions (including stability, protection from ionizing radiation during deposition, and mechanical energy of the environment). Some records of habitability, especially the availability of raw materials, may not be detectable from orbital observations and require in situ observations.

The old sediments enriched in Fe/Mg phyllosilicates and the alteration profiles of the upper stratigraphy containing the most diverse mineralogy on Mars indicate favorable habitable paleoenvironment both in terms of geochemical and depositional conditions. The presence of several hydrated mineral units with well-preserved stratigraphy over several tens of meters with deposition spanning the middle to Late Noachian are evidence for a long-lived source of liquid water, over one or a few long episodes or over several shorter episodes, and availability of diverse energy sources. The reduced ferrous iron layer between the two major units, as well as the areas particularly rich in kaolinite and showing the presence of ferrous iron (outside of the main ferrous iron layer) offer the opportunities to have a reduced environment. In some locations, these reducing phases are in close proximity to oxidized Fe/Al-sulfates, indicating redox gradients at the surface, in the subsurface, or at the interface between the two. These changing redox conditions suggest active chemistry in some strata of the Mawrth Vallis sediments (Bishop et al., 2008). The presence of hydrated, poorly crystalline materials sometimes associated with sulfates at the top of the stratigraphic column likely marks also a change in aqueous environment (possibly related to change in climate). A scenario consistent with the observed mineralogy would involve pedogenic alteration by rain or snow to produce the observed Al-phyllosilicates and ponding of runoff or groundwater to produce sulfate salts. The interaction between reducing groundwater and a more oxidizing surface environment may have created a redox gradient to produce specifically Fe/Al-sulfates (e.g., Hurowitz et al., 2014). Some exposures are visually consistent with oxidation of reducing fluids, e.g., at acid-saline seeps or springs. Ferrous and thus presumably reducing alteration zones are also present at the top of the Fe/Mg unit, consistent with a reducing aquifer perched at the contact. In both these surface and subsurface environments, Fe/S redox gradients may have provided a major energy source. The remnants of ancient valleys and ponding features may have resulted in locally distinct sediment deposits that precipitated in a circum-neutral/low salinity, i.e. habitable environment (Bishop et al., 2018b; Dohm et al., 2011). These sapping or ponding features could be a typical habitable target. All these settings indicate potential habitability in low-energy circulating water and ponded water environment, which may have provided favorable condition for photosynthesis at the surface or chemosynthesis in redox gradients.

Large light-toned veins/ridges, and inverted veins where erosion is stronger, were mineralized by fluid circulation in ancient fractures. Precipitated minerals in these fractures may have preserved and concentrated fluids from subsurface environments. One working hypothesis is that the mineralization occurred under low-T hydrothermal conditions. Hydrothermal deposits represent a category of an ancient geologic environment long recognized as an important target for exploration of Mars due to their habitable potential (Walter & Des Marais, 1993; Farmer & Des Marais, 1999; Michalski et al., 2018, Onstott et al., submitted). Fractures can focus chemical and physical gradients allowing reactants and products to be exchanged over long time scales. Lithological interfaces where zones of focused fluid flow (such as the veins networks observed in Mawrth Vallis) are considered to be excellent targets for exchange of materials with the environment (Onstott et al., submitted). Fracture networks also provided a conduit between habitable subsurface aquifers and more transient surficial habitable systems. Source of hot water is an open key question. Earth-generated hydrothermal environments include deep- and shallow-sea vents (black and white smokers), subaerial hot springs, as well as subaqueous lacustrine spring-vents and hot-water fluvial systems. Except for the first ones (deep-sea vents), we cannot exclude the other ones as analogues for Mawrth Vallis given the geological context of the region.

The large thickness implies rather an extremely long-lived aqueous system (and hence favorable to habitability). Deposition over a significant period of time is also supported by the fact that the layered section contains evidence for buried impact craters and folding or channeling (i.e., evidence for unconformities within the stratigraphic package). In addition, the regional extent of the upper altered sequence is consistent with a sub-aerial weathering origin for the clays, either due to top-down leaching of the stack or weathering concurrent with sedimentation to form a paleosol sequence. These stratigraphic characteristics thus suggest build-up of stratigraphy through discrete depositional events over a duration of time rather than in a single catastrophic event.

## 4.2. Preservation conditions

Several key factors that facilitate the biosignature preservation (e.g. Farmer and des Marais, 1999; Summons et al., 2011; Westall et al., 2015; Hays et al., 2017) can be found in the Mawrth Vallis region (Michalski et al., 2010; Bishop et al., 2013; Horgan et al., 2015):

- Rapid mineral precipitation, which traps and seals organics and textural biosignatures. On Earth, long-term preservation is most successful in host rocks composed of stable minerals that are resistant to weathering and provide an impermeable barrier for the biosignatures. Hydrated silica (especially allophane), sulfates and carbonates can participate to this process at Mawrth. Jarosite is found in association with high organic concentrations in terrestrial paleosols analogous in many ways to that of the Al-phyllosilicate-bearing units (Noe Dobrea et al., 2016). Springs are excellent sites for preservation of morphologic, organic, and other biosignatures (Hays et al., 2017). Other local processes and/or environments within the soils can help to further enhance organic preservation, including reducing paleosols (e.g., wetlands, Hays et al., 2017) and silica deposition, both of which are inferred at Mawrth.

- Rapid burial and recent exhumation. Preservation of biosignatures is favored in rapid burial conditions in fine-grained, clay-rich systems or by chemical precipitation of clay minerals and silica in void space (Farmer and Des Marais, 1999; Bishop et al., 2013). Based on the orbital observations, it is unclear whether the thick (~300 m) stratigraphic package of layered rocks at Mawrth represents rapid deposition. As discussed in the previous section, build-up of the full stratigraphy through discrete depositional events over a duration of time rather than in a single catastrophic event suggests actually slow burial. Only the paleosol sequence could have promoted rapid burial, which protected materials from erosion, oxidation, and degradation at the surface.

At geological timescale, the clay-bearing deposits were rapidly capped by a regionally-extensive mafic 3.6 Gy-old dark mesa-forming unit that preserves the clays and morphologic features for billions years. The material that was buried below the dark cap unit and only recently exhumed (by erosion or by recent small impact craters) is of great interest in terms of preservation index. Because the terrains around Mawrth Vallis are partially eroded, any rover will have access to many well-preserved and recent exposures that represent different environments recorded throughout the volume of the deposit.

- Abundant clay minerals. The occurrence of large amounts of clays and organic materials facilitates the formation of organo-mineral complexes of low permeability that prevents later fluid flow. The region exhibits the highest clay abundance on Mars in multiple environmental contexts (Poulet et al., 2008; Poulet et al., 2014; Poulet et al., 2018). This promotes the impermeability of sediments and protection against later degradation due to diagenetic processes.

- Anoxic conditions that prevent oxidation of organics. The reduced Fe-bearing alteration phases in ferrous unit suggest surface/subsurface saturated and anoxic conditions. Fe-bearing phyllosilicates such as nontronite, Fe-rich montmorillonite, and micas form preferentially in anoxic waters (Harder, 1988) where oxygen is depleted due to reaction with $Fe^{2+}$ released into solution from basaltic rocks (Bishop et al., 2013). Such waters are characteristic of conditions on early Earth (Kasting and Howard, 2006), possibly similar to past Mars. The oxygen-poor atmosphere on the early Earth also supported formation of Fe-rich smectites such as nontronite from basaltic material in sedimentary environments (Harder, 1988). Similar condition may have existed during the formation of Mawrth Vallis sediments. Once the nontronite forms, some oxidation is however necessary to produce $Fe^{3+}$ and stabilize the nontronite (Lonsdale et al., 1980).

- Limited diagenesis that prevents destruction of organics and textures. The perseverance of smectite clay minerals in this region for billions of years implies that their interaction with water has been extremely limited after the minerals formation (Tosca and Knoll, 2009). Also primarily physical weathering produces illites and siliceous sediments, which are not observed. The rocks are layered essentially everywhere they have been observed; they are intact and relatively flat-lying suggesting that their textural properties have been also preserved. Although they have been buried to some depth and later exhumed, they have not been buried very deeply, nor have they been severely deformed by regional stresses, which is consistent with the absence of compositional diagenesis. The region is also surrounded by large impact craters, which could have broken the strata. But preferred orientation or clear density variations with distance to large craters is not observed. Unlike on Earth, where sediments inevitably become diagenetically altered at the scales of hundreds of My, the lack of evidence for illitization and mechanical stresses due to diagenesis that have occurred elsewhere on Mars (Carter et al., 2013) and reduced biosignature preservation potential implies that the original chemical, mineralogical and textural properties of the rocks are likely to reflect ancient geochemical conditions during the time period in which they formed. The only possible exception to this good preservation index could be the lower sequence of the stratigraphy where the densest networks of fractures are observed.

### 4.3. Biological potential

The astrobiological potential of the Mawrth Vallis sediments due to particular features of smectites chemistry has been extensively discussed in Bishop et al. (2013). Basically, these minerals can catalyze chemical reactions by bringing together organic molecules on their surfaces due to the local acidity of their surfaces, the shape specificity of reaction sites, the motion restriction on water molecules, and the binding properties of specific cations. One of them, montmorillonite, has been found to catalyze a number of organic reactions, including the formation of oligomers of ribonucleic acid (Ferris, 2006).

The diverse composition of phyllosilicate outcrops indicates active geochemical environments supportive of specific conditions, that may have even provided energy sources for microbes, and that could in part be a result of the presence of microbes. For instance, the change in iron is is frequently observed. Abrupt changes in chemistry are often indicative of biogeochemical activity on Earth (Bishop et al., 2012). Organics/biosignatures can be also trapped in precipitated minerals (e.g., silica/quartz) and in fluid inclusions associated to mineralized veins (Onstott et al., submitted).

Astrobiological potential of ancient surface and subsurface environments at Mawrth Vallis can be also evaluated from terrestrial analogues. As already discussed, the diverse mineral phases detected in the full clay-bearing sequence suggest a diverse suite of aqueous environments (paleosols, streams, wetlands, springs, marine, hydrothermal sources) powered by surface water/groundwater interactions. On Earth, these environments present highly variable biological potential depending on the environment in which they were formed and then preserved. Hays et al. (2017) reviews in detail the pro and con of biosignature preservation and detection in Mars analog environments including those assumed to be at Mawrth. Some examples are given in the following.

One of the most widespread habitable environments on Earth are soils (Brantley et al., 2006) and the related subaerial environments (Hays et al., 2017). They provide abundant geochemical sources of energy for microbes, and form even under water-limited or snow-dominated climates. More generally, the complex chemistry observed in these so-called critical zones creates redox gradients due to the proximity of the atmosphere, downward flow of precipitation in soils, groundwater fluctuations, evaporation and concentration of brines and spring emergence, in such a way a variety of microbial communities both at the surface (phototrophs) and in the near-surface (chemotrophs) is supported. As stated before, the high clay content of soils enhances preservation, while reducing terrains, like those formed in wetlands, can preserve high concentrations of organics in the form of coal precursors (Horgan, 2014). Examples of Mawrth Vallis analog paleosol sequences on Earth include the John Day Fossil Beds national monument in Oregon, USA (Horgan et al., 2012) and the Painted Desert in Arizona (McKeown et al., 2009b; Noe Dobrea et al., 2009; Perrin et al., 2018). At these sites, high clay content is observed. For example, paleosols in the John Day fossil beds can contain up to 90 wt. % clay and, most likely because of low permeability created by high clay content, are not significantly altered due to burial (Horgan et al., 2012). One frequent criticism of such sites is that finding high concentrations of organics can be challenging. If oxidized paleosols are indeed not inherently good sites of organic preservation, the reduced ones can once again preserve chemical biosignatures. As example, organic carbon is present in the Archean Denison paleosol at abundances of 0.014-0.25% (Gay and Grandstaff, 1980). The organics are concentrated in the upper 2 meters of the soil profile, and are attributed to near-surface microbial communities. The significant energy availability in cold spring also produces abundant biomass (Beaty et al., 2018), and the biosignature preservation in such settings is enhanced by the diversity of mineral deposits (Hays et al. 2017). Studies of recent subaerial terrains have shown high concentrations of organics occur in jarosite nodules (Noe Dobrea et al., 2016). In summary, although highly variable and diverse, subaerial environments at Mawrth are suitable to contain a variety of habitats that could have promoted both chemotrophic and phototrophic microbial communities with a high organic preservation potential. Reduced clay-rich soils (ideally wetlands-type saturated soils) would be of the highest priority for analysis and collection.

In the absence of clearly recognizable geological features in the thick regional clay-bearing sediments, a well-defined terrestrial analogue(s) at Mawrth is difficult to identify. It could be subsurface, subaqueous and/or subaerial environments depending on the stratas. The astrobiological potential of subaerial environment has been previously presented. On Earth the two other ones support a significant amount of biomass with subaqueous ones being more favorable (Hays et al., 2017). In the case of a subaqueous context, the Mawrth geological characteristics as currently observed are more similar to an open system than a close one. Open systems are often nutrient poor with well oxygenated bottom waters, hence they tend not to preserve high abundances of organic matter. However, the sedimentological character and mineralogy of the layered deposits associated with fluvial structures/ponds may have represented advantageous micro-environmental niches during long periods.

The observed mineralized veins/fractures indicate episodes of groundwater flow through sections of phyllosilicate-bearing rocks. These features are highly promising locales for prospecting for subsurface habitat because of the potential for water to persist, the protection provided from inhospitable surface conditions, and the opportunity to have conduit between habitable subsurface aquifers and more transient surficial habitable systems. On Earth, sedimentary aquifers have been found to be significant microbial habitats (Hays et al., 2017). Conditions in subsurface environments can remain stable for very long periods, allowing microorganisms to adapt to the environment and for energy sources to accumulate. Hydrology and enhanced fluid flow along fractures can focus energy and nutrient sources with the capacity for localized sites of heightened biological activity. Cell concentrations on fracture surfaces are often considerably higher than those of the surrounding fluid or matrix (Onstott et al., submitted). If one assumes that mineralization of the fractures were induced by a hydrothermal system, it enhances the opportunities for microorganisms to acquire energy fluids. The fluids indeed can extract elemental nutrients from host rocks via dissolution and chemical alteration to create solutes and mineral assemblages that are stable. Fossilized life is then concentrated in fractures (Onstott et al., submitted). Samples collected in these veins/fractures with geological context at meter scale or below to understand the chemistry, timing, and longevity of groundwater availability would be thus crucial to interpret habitability and to search for biosignatures.

5.  Mission scenario

5.1.     Mars2020-type mission

To illustrate what a surface mission could be at Mawrth Vallis with a Mars2020-type (and also MSL-type mission), a traverse of 5 km is described in this section, designed to visit as soon as possible the science targets with the highest priority within the first 800 m: the different detected hydrated minerals, the clay sequence transition, the most reduced phases, and ridges and halo-bounded fractures (Fig. 20-21, Table 4). Then material recently exhumed close to the igneous capping unit and the possible inverted channel are the next targets, as well as another area where the rover would go through the clay sequence transition again. The proposed traverse avoids potential soft sands as much as is known from HiRISE imagery, as well as rough areas due to fracturing or erosion of the clay-rich layers. Terrains with slopes higher than 12° at a 2 m-scale were also avoided by the proposed traverse.

More specifically, this mission would explore the following way points:

- way points 1, 3, 4 and 5 would give access to the clay-sequence transition; they have different mineralogy identified with CRISM, respectively Fe/Mg-smectites, and Al-clay minerals spectrally similar to montmorillonite, beidellite and kaolinite. Way point 6 is located in a zone where the doublet phase is mapped.

- Way points 2 and 7 are situated on a halo-bounded fracture and on a bright ridge.

- Way points 8 and 9, located < 2.5 km from the landing point, are near the capping unit. Way point 9, located at the top of the capping unit, was selected as being as close as possible to an unmodified capping unit, while way point 8 is on the talus at the base of the capping unit to investigate capping unit debris as well as clay-bearing material recently exhumed due to the erosional retreat of capping unit.

- Way point 10 is in the Fe/Mg-smectites, on a circular layered feature that could correspond to a ghost crater. Way points 11, 13, 14 and 15 go through the clay sequence transition again, with way point 15 finishing on an identified silica-rich outcrop.

- Way point 12 is chosen on a butte that corresponds to the possible inverted channel.

### 5.2. ExoMars

The simulated landing site (22.39°N, -18.44°E) is chosen near the center of 1-σ LPC ellipse (Fig. 22A) in Mawrth Vallis region. The area is characterized by multiple mesas and light-toned deposits (Fig. 22B). As discussed in 3.2.2, the mineralogy is characterized using the texture of the surface and HiRiSE color images. The proposed touchdown point is located on Al-clay unit.

The 1-km exploration zone within Mawrth Vallis landing envelope involves investigation of the top sequence of clay sediments including several transition zones between Al-clay unit and Fe/Mg-clay unit (Fig. 23). The region also contains layered deposits, caprock unit (in-situ or sampling of specific boulders), large-scale fractures with bright mineral infill and dark infills (sand or caprock unit), inverted fractures and craters, desiccation polygons and dunes. Traversing in this region would help addressing the origin of these geologic features and comparing the samples from multiple adjacent stations to provide temporal and compositional context for sedimentary and/or hydrothermal processes in the Mawrth Vallis region.

Traverse including 30 stations were planned for this site. Hazards (e.g. TARs), steep slopes (>15°), loose surface material were avoided during traverse planning according to the engineering requirements of the mission. Traverse has an average slope of 2.5° and a maximum slope of 9° (Stations 10-11). It is approximately 3000 m in length, taking a total of 300 sols to complete, assuming 10 m/sol traverse. Topography profile between the landing site and final station (Station 30) progresses downhill approximately 20 m across the 3000 m path (Fig. 24).

The traverse travels broadly northeast/north, visiting both clay types, the transition zone and various lithological structures (Table 5). To investigate their origin of the sediments, numerous opportunities can be foreseen (Stations LS, 6-11, 16, 21, 24-29) so as to address the origin of deposition and alteration processes. Areas of high bio-preservation potential exist where fine-grained clay-rich sedimentary units record moderate pH and temperatures and/or extreme redox conditions (Section 4.3). These high priority science targets likely lie beneath intact, protective caprock unit (Stations 6, 11, 19) that have been hither to protected from radiolysis. Additionally, widely exposed mineral veins and fractures are associated with clays (Stations 1-5, 9, 13, 15, 17-18, 20, 22-23, 30) recording ancient, possibly later habitable conditions and can determine the tectonic/sedimentary deformation processes. The sand dunes observed at Station 14 can investigate the nature of wind erosion and material transport on the Martian surface.

## 6. Unique and promising site for future in situ explorations

### 6.1. Robotic exploration

Current overarching goals of robotic in situ explorations include characterizing potential habitable sites and searching for signs of past microbial life (main goals of ExoMars) and preparation of the sample return of collected core samples promising in terms of potential for past life (main objective of Mars2020). Table 6 compares a summary of the major characteristics of the selected landing sites with the proposed Mawrth ellipses for the two forthcoming in situ missions. The remarkable diversity of the science targets in comparison to the selected ellipses for Mars2020 and ExoMars confirms the site as an excellent place to explore for evidence of life or habitability during the past history of Mars. Phyllosilicates within well-preserved delta deposits (Goudge et al., 2015) or lacustrine/palustrine environments (Carter et al, 2016; Vasavada et al., 2018; Quantin et al., 2018a) are often considered as exciting astrobiological targets by the Martian scientific community. However, these settings may not represent sustained aqueous activity over geologically timescales sufficient enough to develop its astrobiological potential in the Martian case (Freissinet et al. 2015; Eigenbrode et al., 2018). Furthermore, access to the largest volume of the deposit is limited. In the case of a preserved delta, a surface rover can only access the edges of the deposit. Oxia planum is basically a flat plain with limited access to vertical investigations of the stratigraphy. For the Mawrth Vallis area, we should have a more accessible mineralogical picture of the aqueous history as indicated by the stratigraphy well visible from the orbit. Multiple aqueous mineral assemblages (and not only weathering sequences as often summarized and simplified by the community) are observed throughout the exposed stratigraphic section increasing its uniqueness in terms of potential habitability. Finally the sites undergone various erosional rates leading the potential to have fresh exposures, while the lower exhumation rate for Oxia could be a threat to the preservation of ancient, complex organic matter (Kite and Mayer, 2017). Future exploration of Mawrth Vallis can therefore offer complementary accomplishments to the forthcoming robotic visits of selected landing sites.

### 6.2. **Human exploration**

Mars is the ultimate "horizon goal" for human space flight. Several factors in regards to choosing a landing site suitable for a human-rated mission have to be considered: EDL characteristics, scientific diversity, and in situ resources in terms of water and feedstocks. Dozens of criteria have been proposed during the first landing site/exploration zone workshop for human missions to the surface of Mars held in 2015 (https://www.hou.usra.edu/meetings/explorationzone2015/). It is out of scope here to evaluate each criterion with respect to the Mawrth Vallis exploration zone proposed during the meeting, but a first and quick assessment has been performed by Horgan et al. (2015). In addition to the science diversity addressed in this paper, the Mawrth region represents one of the most valuable sites for exploration-related resources on Mars. Only two extended regions on Mars, namely Mawrth Vallis and Meridiani, have modal mineralogy dominated by hydrated minerals (Riu et al., 2019). This resource is easily accessible as thick hydrated exposures are visible over a very large surface area of several 100x100 km². The fascinating mineral diversity observed in this region should also provide plenty of feedstock chemical resources including Fe, Al, Si.

Finally, a larger exploration zone of the Mawrth Vallis also offers access to several other notable geologic features, including the dichotomy boundary and the outflow channel (Horgan et al., 2015). Within the Mawrth Vallis channel, there are streamlined islands potentially preserving flood deposits, sulfate sediments, fresh craters into northern lowlands materials, and co-occurring large blocks and filled fractures at several scales that may represent deep crust exposed by the outflows. This region would be thus a scientific and ISRU rewarding, and publicly engaging landing site for human exploration.

## 7. Concluding remarks

Mars space exploration has demonstrated that Martian geology and environmental conditions evolved dramatically during the first billion years. This is recorded in the mineralogical diversity placed in various geomorphological contexts (e.g. Murchie et al., 2009). Mawrth Vallis geological context fits this critical Noachian/Phyllosian time period until early Hesperian (including hence part of the Theikian period). As shown in this paper and previous numerous works on the Mawrth area, its most notable features are thick outcrops of phyllosilicates, a large variety of other aqueous materials, mineralized veins/fractures and fluvial features. These very diverse mineralogical and geological characteristics traces multiple aqueous geochemical environments (occurring in the surface and subsurface) and processes (deposition, sedimentation, alteration, erosion, mineralization, fluvial) through time and small regional variations due to localized processes (e.g. ponds, evaporates, mineralized veins, springs). Mawrth Vallis has uniquely recorded these settings with an excellent preservation index over billions years due to the capping unit. Visiting Mawrth Vallis with robotic or human missions would enable investigations of several likely habitable environments, making it a highly desirable landing site for future in situ exploration. The phyllosilicate-rich sediments and mineralized fractures are promising targets in which to search for evidence of prebiotic chemistry and evidence of life. Landing at Mawrth Vallis will also allow investigation of a full range of minerals detected widely across Mars and exciting geological features that will enable tracing the evolution of Mars' ancient environment and climate. The Mawrth region also represents one of the most valuable sites for exploration-related resources (water, Fe/Al-ores) of non-icy terrains of Mars. It is remarkable that in spite of more than several tens of papers published on this region, new findings are still obtained. Although unfortunately non-selected for the forthcoming in situ missions, the investigation of the Mawrth Vallis region shall continue in order to further advance the understanding of this complex and fascinating site and therefore of the planet.

**Acknowledgments**

We thank all people who supported the Mawrth Vallis during the numerous meetings planned for the selection of the Mars2020 and ExoMars landing sites spanning several years. Their participation provided significant material and discussion that have formed the basis of this work. We would like to thank the HRSC team for the provision of mosaic of nadir images and DEMs.

Kasting, J.F., and Howard, M.T. (2006) Atmospheric composition and climate on the early Earth. *Philos. Trans. R. Soc. Lond.*, *B, Biol. Sci.* 361:1733–1742.

Kinzelman, P. and Horgan, B. (2019) Preservation of surface and subsurface on Mars in filled fractures at Mawrth Vallis. *Journal of Purdue Undergraduate Research* 9:42-48.

Kite, E.S., and Mayer, D.P. (2017) Mars sedimentary rock erosion rates constrained using crater counts, with applications to organic-matter preservation and to the global dust cycle. *Icarus* 286:212-222.

Kocurek, G., and Hunter R.E. (1986) Origin of Polygonal Fractures in Sand, Uppermost Navajo and Page Sandstones, Page, Arizona. *SEPM Journal of Sedimentary Research* 56 doi:10.1306/212f8a7b-2b24-11d7-8648000102c1865d

Lonsdale, P.F., Bischoff, J.L., Burns, V.M., Kastner, M., and Sweeney, R.E. (1980) A high-temperature hydrothermal deposit on the sea bed at a Gulf of California spreading center. *Earth and Planet. Sci. Let.* 49:8–20.

Loizeau, D., Mangold, N., Poulet, F., Bibring, J.-P., Gendrin, A., Ansan, V., Gomez, C., Gondet, B., Langevin, Y., Masson, P., and Neukum, G. (2007) Phyllosilicates in the Mawrth Vallis region of Mars. *J. Geophys. Res.* 112:1–20.

Loizeau, D., Mangold, N., Poulet, F., Ansan, V., Hauber, E., Bibring, J.P., Gondet, B., Langevin, Y., Masson, P., and Neukum, G. (2010) Stratigraphy in the Mawrth Vallis region through OMEGA, HRSC color imagery, and DTM. *Icarus* 205:396–418.

Loizeau, D., Werner, S.C., Mangold, N., Bibring, J.-P., and Vago, J.L. (2012) Chronology of deposition and alteration in the Mawrth Vallis region, Mars. *Planetary & Space Sci.* 72:31–43.

Loizeau, D., Mangold, N., Poulet, F., Bibring, J.P., Bishop, J. L., Michalski, J., and Quantin, C. (2015) History of the clay-rich unit at Mawrth Vallis, Mars: High-resolution mapping of a candidate landing site. *J. Geophys. Res.* 120:1820–1846.

McMahon, S., van Smeerdijk Hood, A., and McIlroy D. (2016) The origin and occurrence of subaqueous sedimentary cracks. *Geological Society, London, Special Publications* 448(1), 285–309.

Malin, M.C., and Edgett, K.S. (2000) Sedimentary rocks of early Mars. *Science* 290:1927–1937.

Mangold, N., Loizeau, D., Gaudin, A., Ansan, V., Michalski, J.R., Poulet, F., and Bibring, J.P. (2010) Connecting Fluvial Landforms and the Stratigraphy of Mawrth Vallis Phyllosilicates: Implications for Chronology and Alteration Processes [abstract 1547]. In *1st International Conference on Mars Sedimentology and Stratigraphy*, 40.

McKeown, N.K., Bishop, J.L., Dobrea, E.Z.N., Ehlmann, B.L., Parente, M., Mustard, J.F., Murchie, S.L., Swayze, G.A., Bibring, J.-P., and Silver, E.A. (2009a) Characterization of
60

| Regional mineralogy | Minerals in ExoMars ellipse | Minerals in Mars2020 ellipse |
|---|---|---|
| $Fe^{3+}$/Mg-smectites | $Fe^{3+}$/Mg-smectites | $Fe^{3+}$/Mg-smectites |
| $Fe^{2+}$/Mg-clays | $Fe^{2+}$/Mg-clays | $Fe^{2+}$/Mg-clays |
| $Fe^{2+},Fe^{3+}$,Al,Mg vermiculitic | $Fe^{2+},Fe^{3+}$,Al,Mg vermiculitic | ? |
| Al-hydrated phases (kaolinite, montmorillonite, beidilite) | Al-hydrated phases (kaolinite, montmorillonite?, beidilite) | Al-hydrated phases (kaolinite, montmorillonite, beidilite?) |
| Si-OH phases | Si-OH phases | Si-OH phases |
| Zeolites | ? | ? |
| Ferrihydrite, NpFeOx | Ferrihydrite, NpFeOx | Ferrihydrite, NpFeOx |
| Allophane, Imogolite | Allophane | Allophane |
| Sulfates (bassanite, jarosite, alunite) | Sulfates (jarosite, alunite) | Sulfates (jarosite, alunite) |
| Carbonates | ? | ? |
| Mica (celadonite) | ? | Mica (celadonite) |

TABLE 1. Mineral identification in the Mawrth regions and the proposed ellipses.

| Science targets | Zone 1 | Zone 2 | Zone 3 | Zone 4 |
|---|---|---|---|---|
| Sediments enriched in phyllosilicates | Al/Si and Fe/Mg | Al/Si and Fe/Mg | not at surface | Al/Si and Fe/Mg |
| Upper stratigraphy of well-preserved profiles | confirmed incl. kaolinites | confirmed incl. kaolinites | suspected right under cap rock 2-4.5 km | confirmed |
| Veins | present | present | assumed | present |
| Ridges | present | no | no | present |
| Fractures | present | present | certain < 4.5 km | present |
| Fresh material under the capping unit | accessible < 1 km | accessible < 1 km | accessible through erosive windows | accessible < 1 km |
| Fluvial and ponding features | inverted channels | filled channel within 2 km EZ | no | filled channel within 1 km EZ |
| Sulfates deposits | assumed | confirmed < 1 km | no | confirmed < 4.5 km |

TABLE 2. Science target diversity and accessibility within four examples of exploration zones (EZ hereafter) within the ExoMars landing envelope. Images are shown in Fig. 17.

| Science targets | Habitable potential | Potential for biosignatures | Preservation properties |
|---|---|---|---|
| Sediments enriched in Fe/Mg phyllosilicates | High: long and sustained aqueous activities with multiple possible subaqueous, subaerial and/or subsurface settings (marine-pedogenesis-sedimentary-fluvial-transport), availability of diverse energy sources due to geochemical diversity | Smectites-bearing rocks as reactants for pre-biotic chemistry and/or habitats for microbes | Burial processes in clay-rich layered systems. Abundant clay concentration. No mineralogical and lithological evidence for diagenesis |
| Upper stratigraphy of well-preserved profiles | High: very diverse mineralogy, multiple possible aqueous subaerial settings (pedogenesis-palustrine-wetland-spring), availability of energy sources due to various geochemical conditions, proximity for photosynthesis | Redox gradients as biological indicator, phyllosilicate-bearing rocks as reactants for pre-biotic chemistry | Anoxic conditions. Abundant clay concentration. Sealing by amorphous silica. Buried by the capping unit |
| Large light-toned veins, ridges, inverted veins | High: fluid circulation in ancient fractures enhancing nutrients concentration, possible low-T hydrothermal activity, conduit between potentially habitable subsurface and surface sites, lithological interfaces | Potentially higher biomass, redox and lithological interfaces as biological indicators | Buried by the capping unit, more indurated material than the sedimentary one |
| Fluvial features | Medium: low-energy circulating water and ponded water | Local concentrations in | Buried by the capping unit, local |

|  | environment | sediments | fine-grain sediments |
|---|---|---|---|
| Sulfate deposits | High: strong redox gradient, wetland-acid/saline seep-spring environments | Help biomass concentration, Redox gradients as biological indicator | Buried by the capping unit |

TABLE 3. Potential for habitability, preservation and biosignatures of the major science targets of Mawrth Vallis deposits. See text in section 4 for details.

| Science targets | Station number | Major science rationale |
|---|---|---|
| Fe/Mg clay unit | 1, 2, 6, 7, 10, 11 | Access to moderate W/R (both subsurface and surface) environments |
| Al/Si/allophane/(carbonate) unit | 3, 4, 5, 13, 14, 15 | Access to subaerial environments |
| Clay-sequence transition | 1-5, 10-15 | Access to a gradient of ancient environmental conditions, and to the chronology of depositional conditions |
| Reduced environments | 5, 14, 15 | Possibly the best preservation conditions for some biosignatures |
| Ridges and halo-bounded fractures | 2, 7 | Subsurface aqueous activity including potential hydrothermal environment |
| Large-scale fractures with dark infills | 2, 6, 7, 10, 11, 13, 15 | Evolution of the clay-rich unit under diagenesis and atmospheric constraints |
| Clay-bearing material | 8 | Material very recently exhumed due to the erosional retreat of capping unit |
| Capping unit | 9 | Igneous history of the planet and datable surface |
| Ghost crater | 10 | Access to a different depositional environment |
| Desiccation polygons | 1-7, 10-15 | History of the Martian atmosphere and evolution of the clay-rich unit |

| Dunes, debris | 2, 5, 8 | Recent history of the Martian atmosphere of erosional processes |
| Possible inverted channel | 12 | Fluvial activity, possible deposition of transported material |

TABLE 4. List of way points for each science target for a Mars2020-type mission.

| Science targets | Station number | Major science rationale |
|---|---|---|
| Fe/Mg clay unit | 7, 8, 16, 21, 24, 25, 29, 30 | Access to moderate W/R (both surface and subsurface) environments and transition zones |
| Al/Si/allophane/(carbonate) unit | LS, 6, 9-15, 19, 20, 26-28 | Access to subaerial environments |
| Layered materials | 10-11, 26-28 | Access to changing sedimentary and chemical conditions |
| Large-scale fractures with bright mineral infills | 1-3, 5, 13, 15, 17, 18, 20, 22, 23, 30 | Subsurface aqueous activity including possible hydrothermal environment |
| Large-scale fractures with dark infills | 9 | Evolution of the clay-rich unit under diagenesis and atmospheric constraints |
| Inverted fractures (or Positive-relief ridges) | 4 | Differential erosion, preserved fluid pathway |
| Capping unit | 6, 11, 12, 19 | Igneous history and datable surface. Highest preservation potential of biosignatures underneath. |
| Desiccation polygons | Potential in 6, 9-15, 19, 20, 26-28 | History of the Martian atmosphere and evolution of the clay-rich unit |

| Dunes | 14 | Modern erosional history |

TABLE 5. List of the proposed science targets of the ExoMars mission scenario. Examples of these targets are shown in Fig. 25.

| Criterion | Jezero crater | Oxia Planum | Mawrth Vallis | |
| --- | --- | --- | --- | --- |
| | | | Mars2020 ellipse | ExoMars ellipse |
| Timing of aqueous activities | Late Noachian/Early Hesperian | Mid-Noachian | Late Noachian to early Hesperian | Mid-Noachian to early Hesperian |
| Environment settings of aqueous sediments | Sedimentary, fluvio-deltaic, lacustrine | Lacustrine, palustrine | Sedimentary, pedogenetic, palustrine, springs, low-T hydrothermal | Sedimentary, pedogenetic, palustrine, springs, low-T hydrothermal |
| Aqueous geochemical environments indicated by mineral assemblages | Fe/Mg smectites, carbonates, Fe-hydroxide | $Fe^{3+}$/Mg vermiculitic phases, Fe-hydroxide | $Fe^{3+}$/$Fe^{2+}$/Mg smectites, Al-phyllosilicates, sulfates, allophanic material, Si-OH phases, Fe-hydroxide, NpOx, Mica (celadonite), carbonates? $Fe^{3+}$/Mg vermiculitic phases? | $Fe^{3+}$/$Fe^{2+}$/Mg smectites, Al-phyllosilicates, sulfates, allophanic material, Si-OH phases, Fe-hydroxide, NpOx, Mica (celadonite)?, carbonates?, $Fe^{3+}$/Mg vermiculitic phases |
| Freshness of the surface | Continuous erosion of | Continuous erosion of | Very fresh exposure behind | Very fresh exposure behind eroded |

|  | deltaic deposits | hydrated sediments | eroded capping unit | capping unit |
|---|---|---|---|---|
| Preservation from radiation after deposition | Covered by dark-toned, mafic unit (~2.6 Gy) | Exposed during ~1Gy as oldest capping (lava flows) is 2.6 Gy Strong erosion (900m?) | Covered by capping unit (3.6 Gy) | Covered by capping unit (3.6 Gy) |
| Diagenesis/metamorphism | None | Possible hydraulic fracturing | Possible fracturing due to Oyama impact | Possible hydraulic fracturing for the lowest part of the sediments |
| Evolution of the climate | Yes | Yes | Yes (over a larger period of time) | Yes (over a larger period of time) |
| Potential for biosignatures | Deltaic deposits, lacustrine basin fill (clays, carbonates) | Sediment layers | Sediment layers, ponds, mineralized fractures, springs, paleosols | Sediment layers, ponds, mineralized fractures, springs, paleosols |
| Water resources (based on hydrated minerals %Vol) | Lower than the other sites | High | Highest | High |

TABLE 6. Comparison of selected properties between the different sites. Sources are various: Goudge et al. (2015; 2018; personal communication), Shahrzad et al. (2019) for Jezero, Quantin-Nataf et al., (2018a; 2018b) for Oxia. Water resource evaluation come from Poulet et al. (2014; 2018a; 2018b) and Riu et al. (2019).

**FIG. 1.** The Mawrth Vallis region with the proposed landing site areas over (left) MOLA altimetry superposed on the THEMIS (Thermal Emission Imaging System) daytime IR mosaic and (right) HRSC (High Resolution Stereo Camera) color and nadir mosaic. The ellipse envelopes are drawn in white for Mars2020 (12 km x 10 km) and red for ExoMars. The ellipses (120 km x 19 km) for ExoMars Launch Program Opening and Closing (LPO and LPC respectively) are in yellow.

**FIG. 2.** Schematic stratigraphy of the Mawrth Vallis layered deposits compared to the vertical distribution of the altered phases. The dark cap overlays (and protects) the two main compositional zones dominated by Al/Si-OH phases in cyan and by Fe/Mg smectites in red. Ferrous material is drawn at the transition of the two main units as green dashed zone. Fractures are represented by crossed lines or filled lines when fluid circulation occurred (see text). Sulfates were detected in the lower section of the strata (bassanite, blue layer), but also as localized deposits (identified as jarosite, alunite but not represented here). It has been suggested that the lower strata where the densest network of fractures is visible could correspond to the deposits of Oxia Planum (section 2.3)

**FIG. 3.** Fluvial structures around Mawrth Vallis channel. A) shows in light blue **lines** all main identified valleys and inverted valleys around the region, on the plateaus, the floor of Mawrth Vallis and the floor of Oyama Crater. The background image in an HRSC nadir image mosaic, with superimposed HRSC DTM mosaic (largest ejecta blankets mapped in gray). B) and C) display along dashed lines examples close to the proposed landing ellipses on an HRSC image mosaic, with difference in elevation of 1000 m in both images. In (D) a large ponding geological feature in the eastern part of the ExoMars landing site is observed (white rectangle).

**FIG. 4.** Examples of the varying fracture patterns found in the Mawrth Vallis region on HiRISE (High Resolution Imaging Science Experiment) color images. A and B come from the vicinity of the Mars2020 landing ellipse (respectively ESP_020798_2040 and ESP_009115_2040), C and D are found within the ExoMars landing ellipse (ESP_045114_2025 for both). (A) shows a relief where top is north and bottom is south, the small polygonal texture (similar to elephant skin) is present at the top of the stratigraphy and the densest set of irregular fractures is lower.

**FIG. 5.** Schematic depositional history of the altered and layered deposits at Mawrth Vallis. A: Progressive deposition and alteration of clay sediments. B: Progressive deposition and alteration of clay sediments and hydraulic fracturing due diagenesis. C: Deposition of top section of Al-clays/Si-OH/Allophane/Ferrous clays/Carbonates, fluid circulation in fractures, fluvial activity. D. Deposition of sulfates. E. Deposition of the capping unit.

**FIG. 6.** Location of the proposed Mars2020 ellipse, on a plateau between Mawrth Vallis and Oyama crater, centered -19.06°E, 23.97°N. The figure shows HiRISE images over a HRSC nadir images mosaic, with a HRSC DEM mosaic superimposed in color.

**FIG. 7.** Location of the proposed Exomars landing ellipses (green) with the according envelope (red) over HRSC color image mosaic background.

**FIG. 8.** Map of geologic features within the landing ellipse defined for the Mars2020 mission (background is made of HiRISE ortho images DTE1A_010816_2040_010882_2040, DTE1A_008469_2040_008825_2040, DTE1A_005964_2045_011884_2045). Nearly the whole map appears light-gray as the site was chosen to maximize clay surface for direct landing on the clay unit.

**FIG. 9.** Examples from HiRISE images of geological features mapped in Fig. 8 within the landing ellipse proposed for Mars2020. C-E are from HiRISE color observations. I has superimposed HiRISE DEM with a 10 m maximum difference in elevation. (HiRISE image IDs: A: PSP_008469_2040, B: PSP_008469_2040, C: PSP_008825_2040, D: PSP_008825_2040; E: PSP_010882_2040; F: PSP_008469_2040; G: PSP_008469_2040; H: PSP_008469_2040; I: DTEEC_010816_2040_010882_2040 with DT1EA_010816_2040_010882_2040; J: PSP_008469_2040; K: PSP_010816_2040).

**FIG. 10.** Map of main geologic units within the landing envelope defined for the ExoMars mission (Context Camera CTX image mosaic). Yellow circles refer to examples of exploration zones described in section 3.4.2.

**FIG. 11.** Morphology of the light-toned terrains observed in the southeastern part of the envelope. Different textures and colors are observed (B), which are typical of the compositional sub-units as reported in many studies on the Mawrth Vallis region. Layers are also visible on (B) and on area where the clay-bearing stratigraphy is eroded (C). Remnant dark knobs of capping unit in the lower right of (A) and (B) are present. Dark toned dust and sand forms small dune fields around these knobs.

**FIG. 12.** Erosional windows in the nort-western part of the ExoMars landing site (green boxes: craters, white ovals: eroded terrains) within the capping unit revealing light-toned materials. The deposits buried below the cap unit were only recently exhumed suggesting an excellent potential in terms of preservation of potential organics. The yellow line defines the boundary of the LPC ExoMars ellipse.

**FIG. 13.** Selected VNIR spectra of alteration phases observed in the ellipse (from Bishop et al., 2016b, 2018a). A) Al- and Si-rich phases in CRISM image FRT0000C467 compared with lab

reflectance spectra of minerals. Spectrum 1 contains features characteristic of the Al-sulfate alunite, spectrum 2 contains features consistent with the Al-rich phyllosilicate halloysite/beidellite, spectrum 3 is similar to the Al-rich phyllosilicate montmorillonite, and spectrum 4 is more similar to opal. B) Fe-rich materials in CRISM image FRT0003BFB compared with lab reflectance spectra of minerals. Spectrum 5 contains features similar to jarosite, spectrum 6 exhibits the "doublet type" feature between 2.2-2.3 µm, and spectrum 7 is characteristic of Fe/Mg-smectite or Mg-bearing nontronite.

**FIG. 14.** Mineral indices from CRISM observation FRT0000B141_07_IF167L_TRR3 that covers almost completely the landing ellipse proposed for Mars2020. Mineral indices are listed as inferred vertical sequence order from alunite near the top to Fe/Mg-smectites at the bottom of the observed sequence. Spectral indices here are based on Viviano et al. (2014) CRISM spectral parameters, and are stacked in the following order. "Alunite" highlights locations where BD2165 > 0.003, BD2165 > BD2190, and MIN2200 < 0.003. "Al-OH (beidellite type)" highlights locations where BD2190 > 0.003 and BD2165 and BD2210_2, and MIN2200 and MIN2250 < 0.003. "Al-OH (montmorillonite type)" highlights locations where BD2210_2 > 0.003 and BD2190, and MIN2200 and MIN2250 < 0.003. "Kaolinite" highlights locations where BD2210 and MIN2200 > 0.003. "Si-OH" highlights locations where BD2190 and BD2210 and MIN2250 > 0.003. "Doublet phase" highlights locations where MIN2250 and D2300 > 0.003, and MIN2250 > BD2190. "Fe/Mg-smectite" highlights locations where D2300 > 0.005.

**FIG. 15.** Mineral map derived from OMEGA and CRISM MSP data overlayed on HRSC nadir image. The mineral phases are similar to the ones found in the Mars2020 ellipses (Figures 13 and 14). Note the small impact crater in the northwest (cut by the green ellipse) exhumes Kaolins and Fe/Mg-rich material. The identification and mapping have been performed according to the methodology defined in Carter et al. (2013; 2018).

**FIG. 16.** Main science targets for the Mars2020 landing ellipse with respect to the mission's scientific goals. The mapping of the mineralogical targets is extracted from Fig. 14. Blank parts are not considered as most desired targets, but as can be seen in Fig. 14, almost the whole surface of the ellipse exposes the clay unit and could be considered a science target as such.

**FIG. 17.** Examples of exploration zones. Their locations inside the ExoMars envelope are indicated on Fig. 10. Images 1-4 derive from CTX and HiRISE with CRISM spectral maps overlain. Color Code: Fe/Mg-smectites (red), Al-phyllosilicates/Si-OH/allophane (blue), Kaolins (turquoise), Sulfates (green). The inset circles define rover exploration zones of 1 km (orange), 2 km (green), 4.5 km (red), namely 2, 4 and 9 km in diameter. EZ1: Located on the light-toned unit surrounded by the knobby cap rock unit. Fractures/veins present. All primary targets located within a 1 km radius. EZ2: Center located on sulfate-rich deposits, transition-zone of phyllosilicates is present within a 1 km radius. High surface coverage with cap rock. Light-toned fractures/veins present. EZ3: Located directly on cap-rock. No phyllosilicate deposits detected from orbit but sporadic light-toned outcrops pointing to the presence of phyllosilicates in minor quantities compared to other EZs. Fractures/veins present at erosional windows. EZ4: Center located on a thinner/eroded capping unit. All phyllosilicate-types as well as fractures/veins present, plus a large outcrop of stratified material.

**FIG. 18.** CTX mosaic with landing envelope. Green regions show the areas where a touchdown point is never more than 3 km away from a prime target.

**FIG. 19.** Conceptual view of the astrobiological relevance of a science target used for this work. It will be the highest if the three conditions for habitability, preservation and potential for biochemical signatures are met.

**FIG. 20.** Traverse (5 km in length) for a mission scenario in a Mars2020-type landing ellipse. The landing point corresponds to the center of the ellipse. The traverse avoids zones with potential soft sand, aeolian ripples, and rough areas due to fracturing and erosion. Red circles give an idea of the distance from the landing point. The white box indicates the position of the close-up in the next slide.

**FIG. 21.** Close-up on the traverse, superimposed on the science targets map at the top, with indication of steep slopes, and superimposed on the CRISM map from Fig. 14 (middle). The traverse profile at the bottom indicates the location of some of the way points. Background image is HiRISE ortho image DTE1A_008469_2040_008825_2040.

**FIG. 22.** (A) The landing site (22.39°N, -18.44°E) is close to the center of 1-sigma ellipse. Landing envelope (green ellipsoid), high-probability landing ellipses for early and late launch dates (green dots), landing site (orange star), exploration zones (1 km - yellow, 2 km – white, and 5 km – red circles) overlaid on MEx HRSC mosaic 50 m/pix. (B) The landing site is on Al/Si-OH phase unit. Area characterized by multiple mesas and light-toned deposits. Primary Mission within 1 km is defined as the exploration zone (EZ) (yellow circle). Base map is MRO CTX 6 m/pix image (D22_035804_2025_XN_22N018W).

**FIG. 23.** A studied landing site scenario and proposed traverse with 30 stations within <1 km EZ overlaid on HiRISE 0.5 m/pix color image (ESP_036305_2025). Zooms over the traverse (black solid line) between stations (white filled circles) 1-20 (top) and 20-30 (bottom). Identification of major targets is schematically indicated. Yellow arrows indicate fractures/veins/ridges. Capping units correspond to the dark mesas. The diversity is more easily observed at higher spatial resolution (Fig. 25).

**FIG. 24.** (Top-left) Slope base map created from HiRISE 1 m/pix DEM overlaid on a MRO HiRISE 0.5 m/pix image (ESP_036305_2025). (Bottom-left) Colorized 1 m/pix HiRISE DEM over the simulated traverse including the 30 stations. (Right) Topographic profile of the simulated traverse with stations progressing downhill approximately 20 m.

**FIG. 25.** Examples from HiRISE images of geological features within the landing ellipse proposed for ExoMars and listed in Table 5. Image subsets of HiRISE 0.5 m/pix color image (ESP_036305_2025).

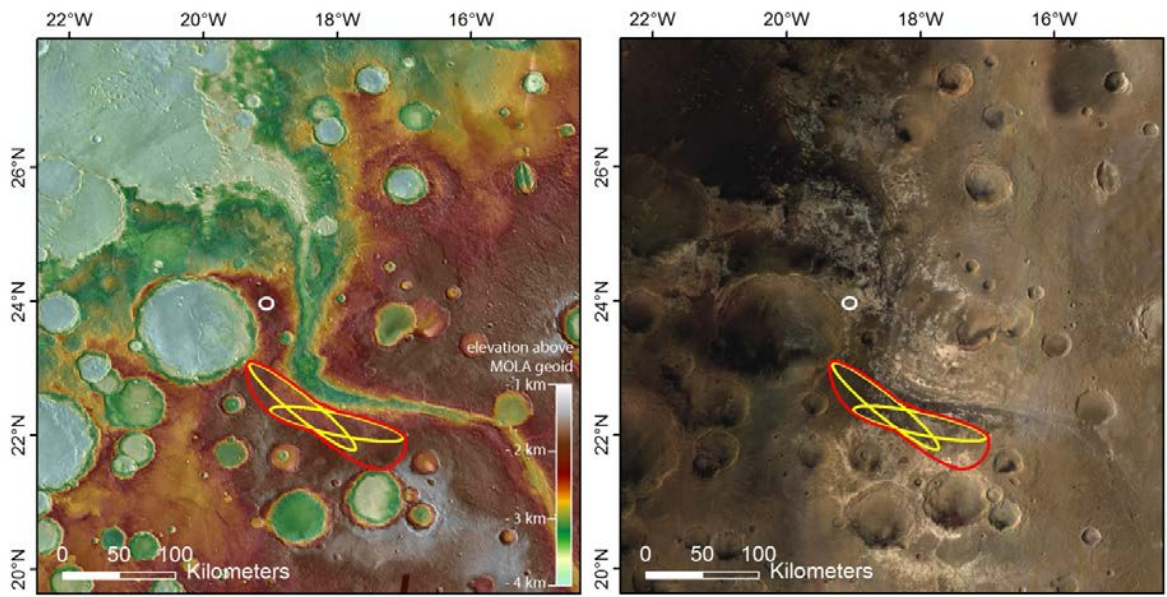

**FIG. 1**

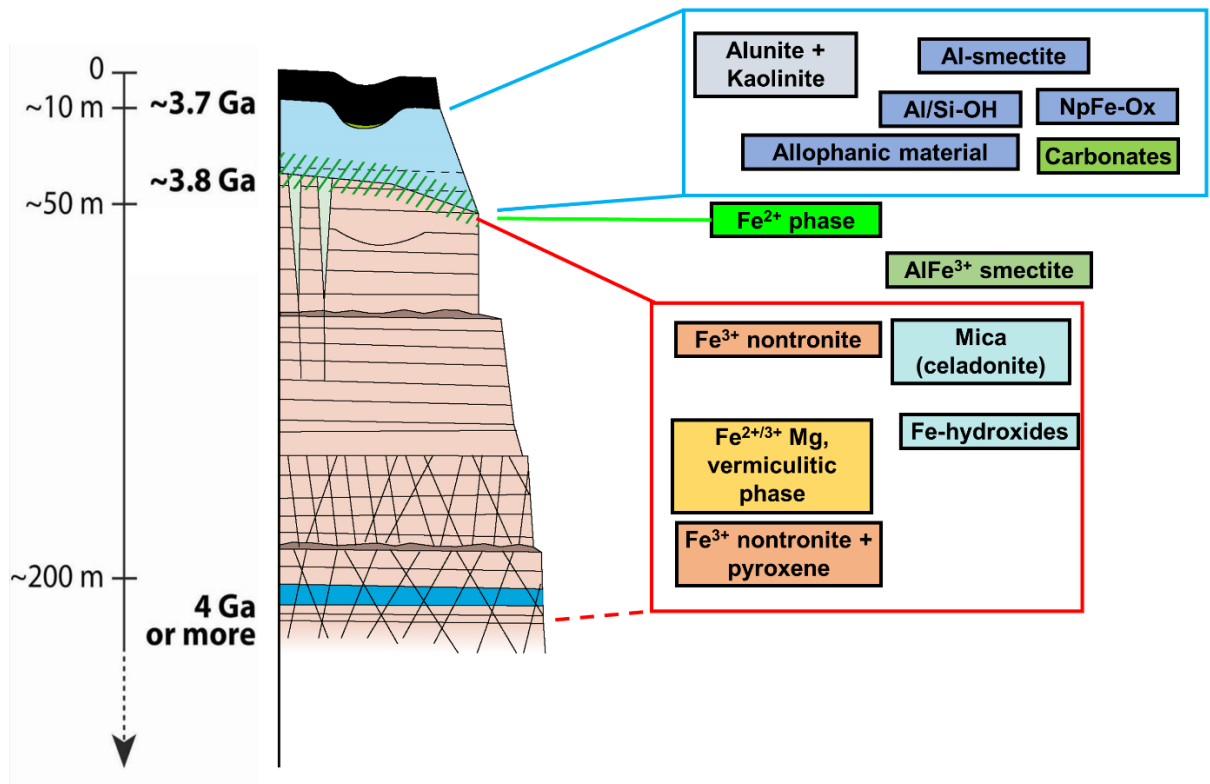

**FIG. 2**

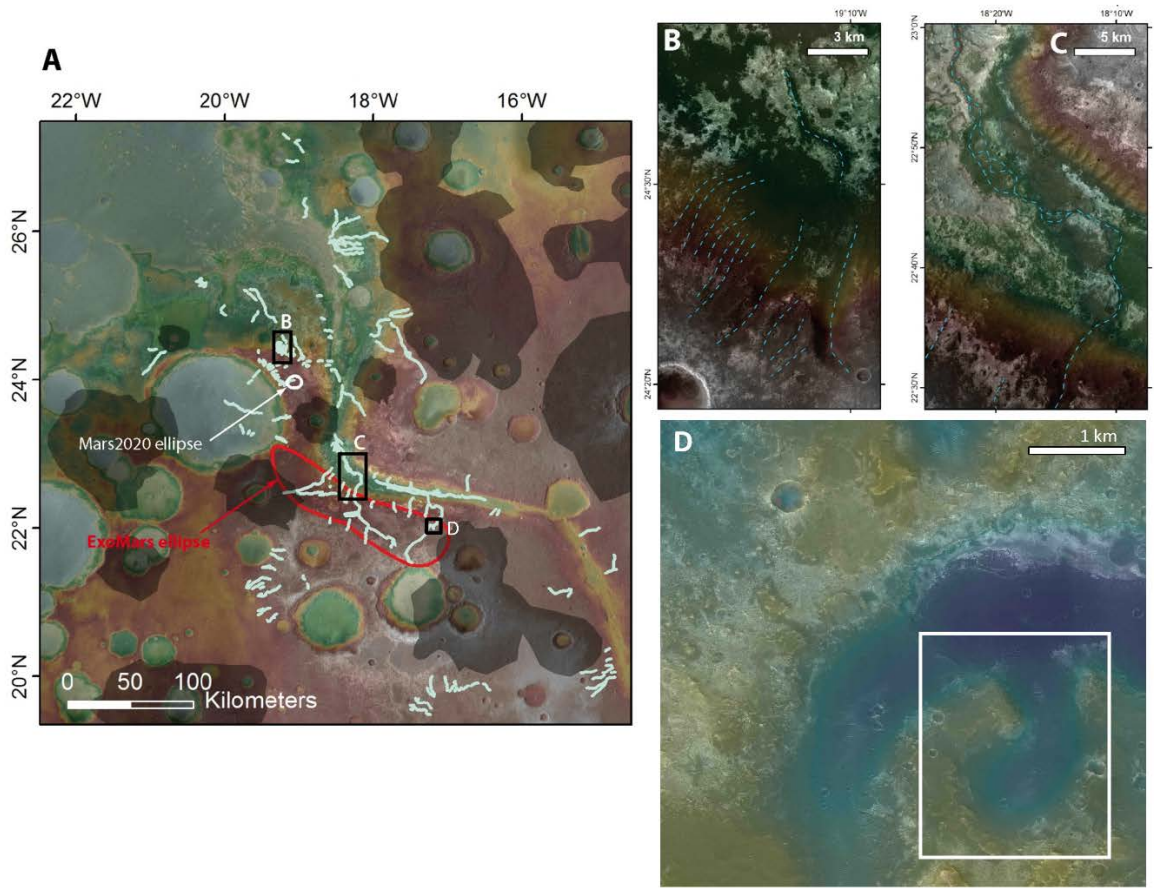

**FIG. 3**

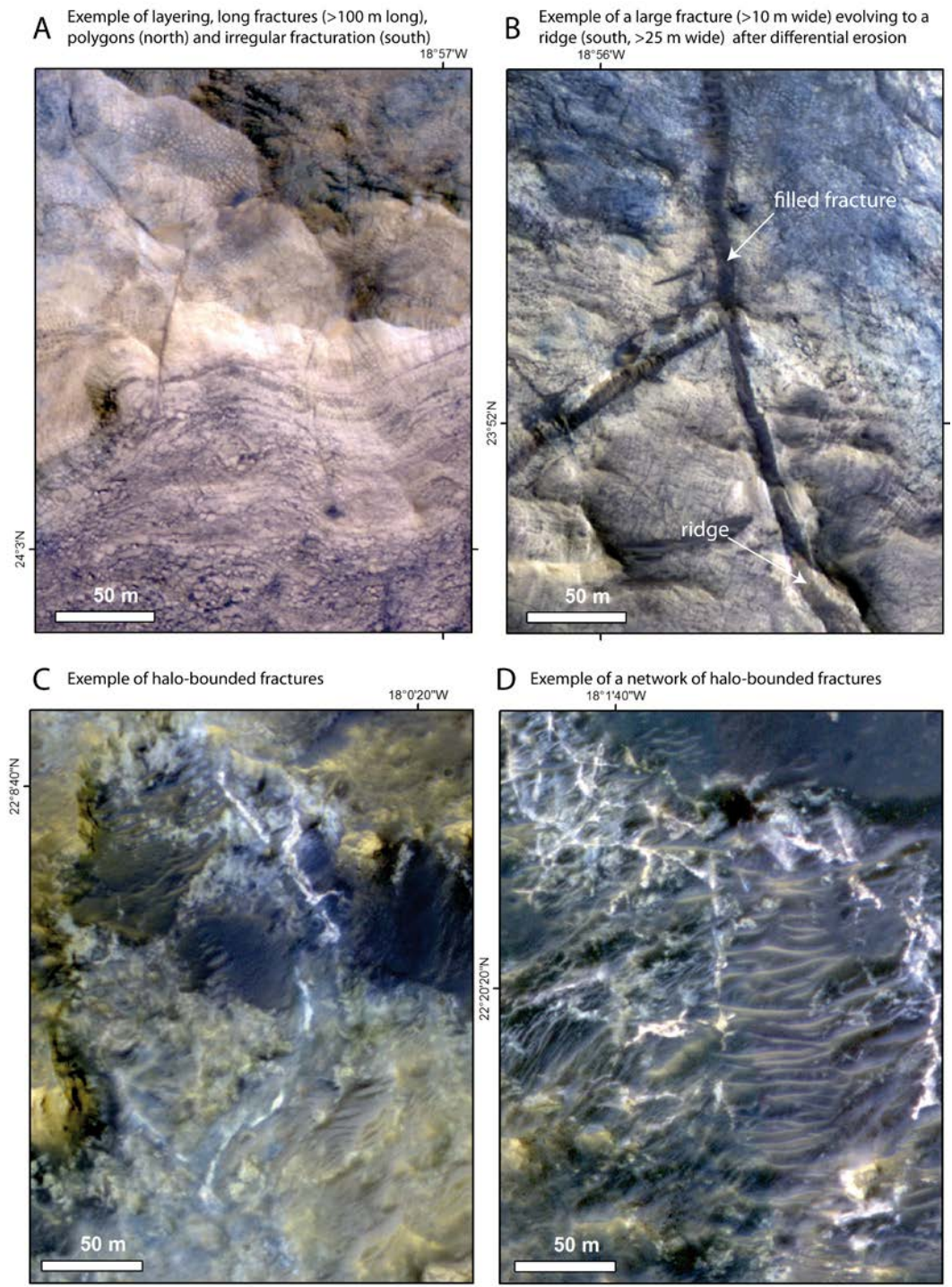

**FIG. 4**

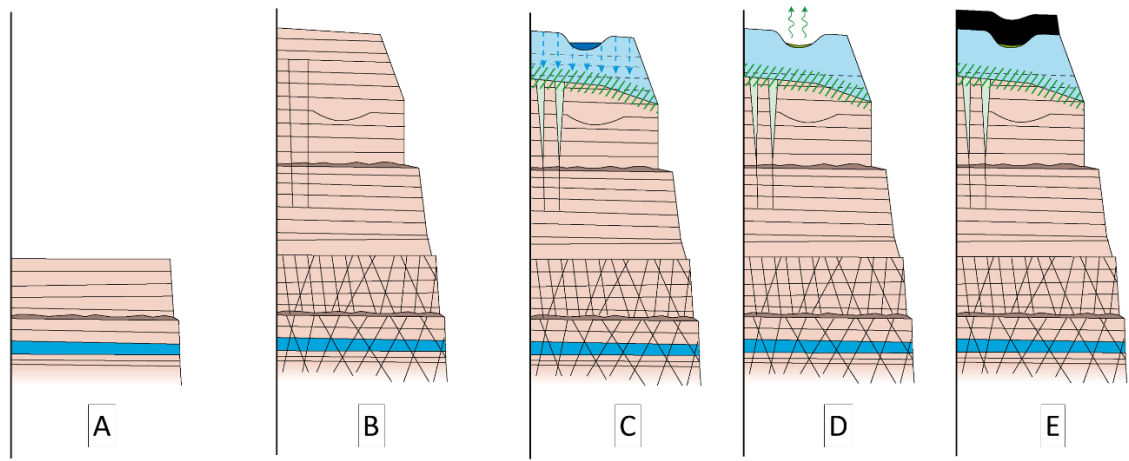

**FIG. 5**

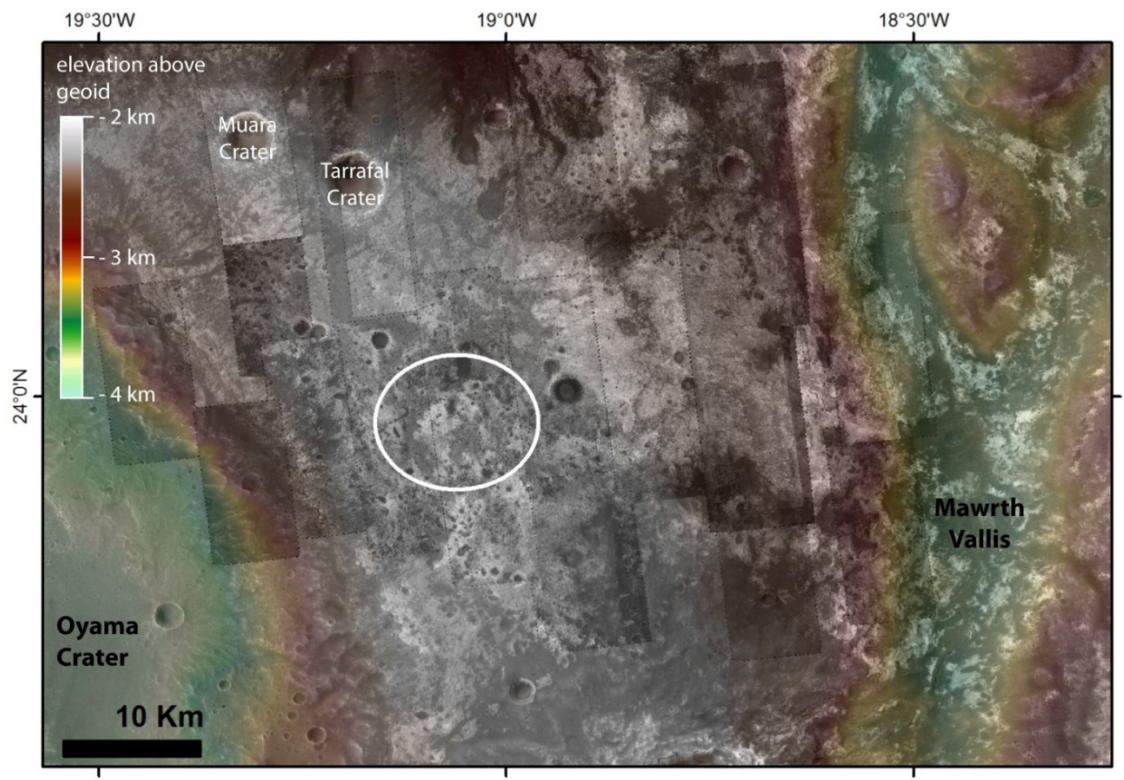

**FIG. 6**

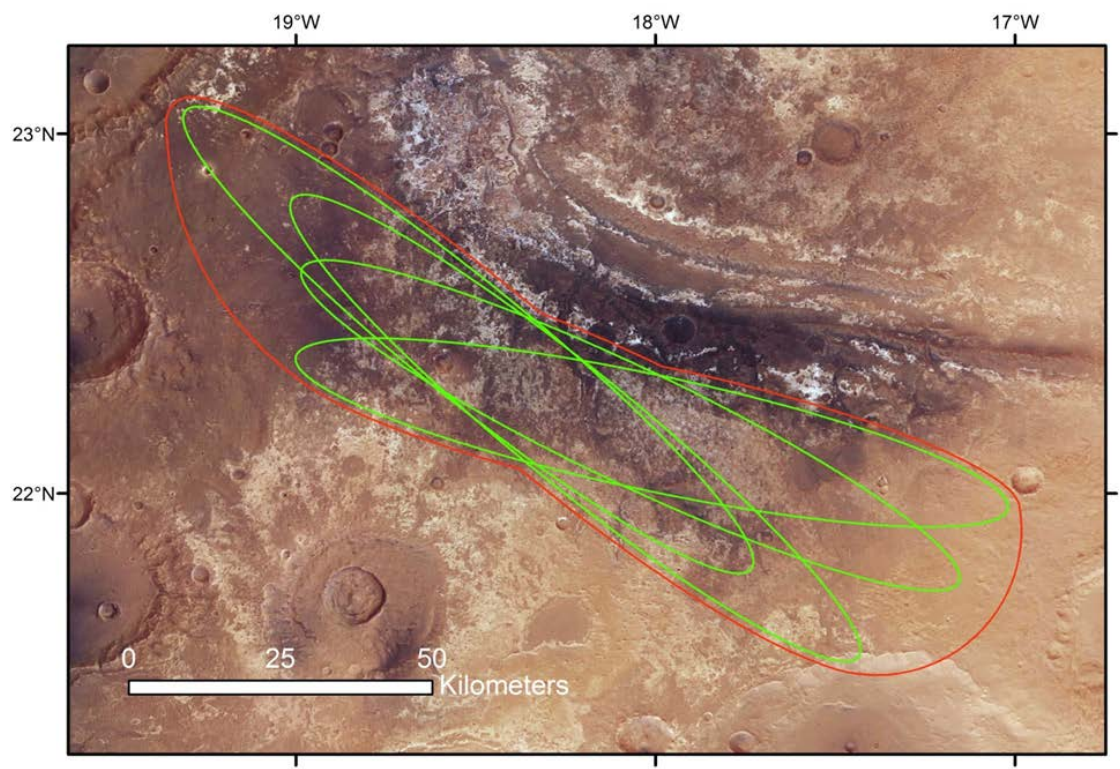

**FIG. 7**

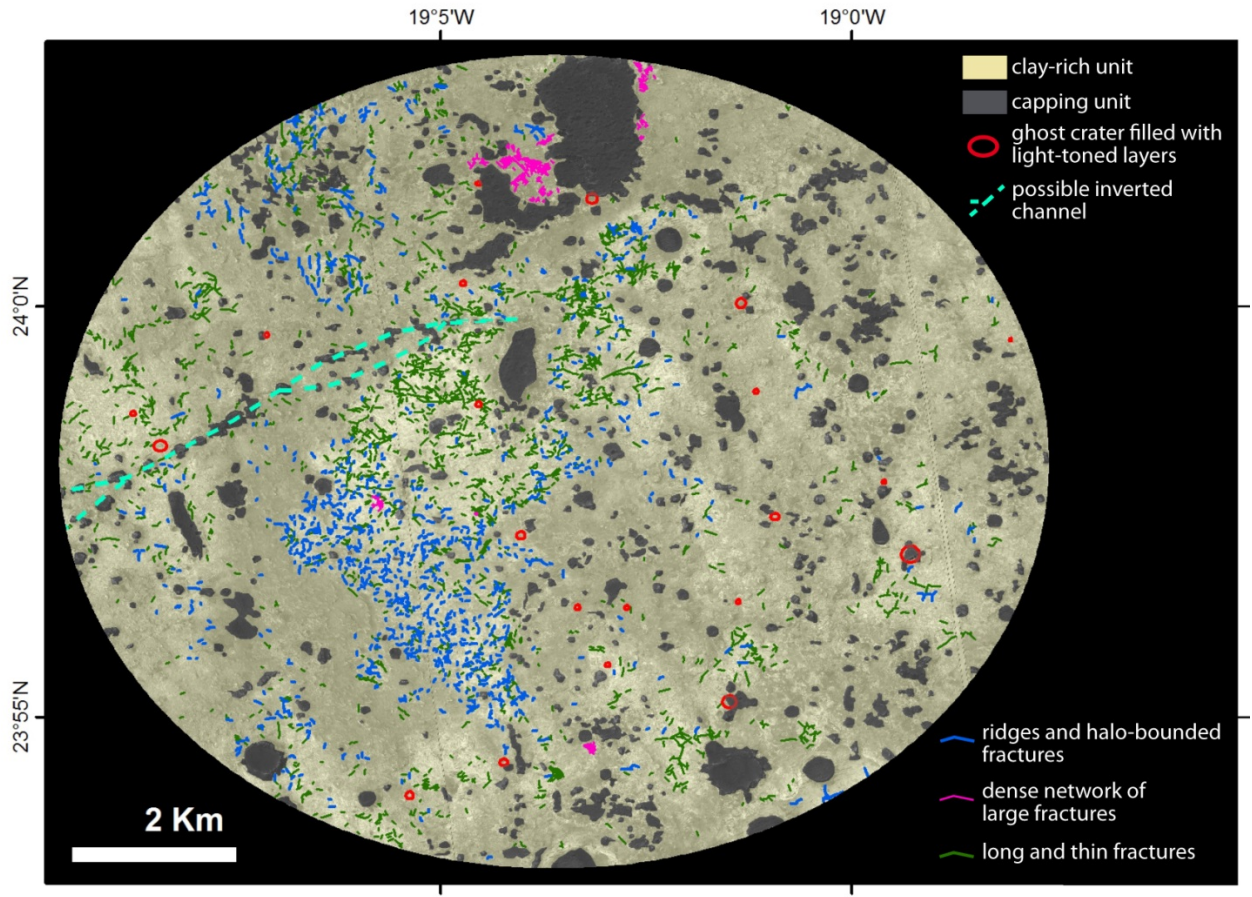

**FIG. 8**

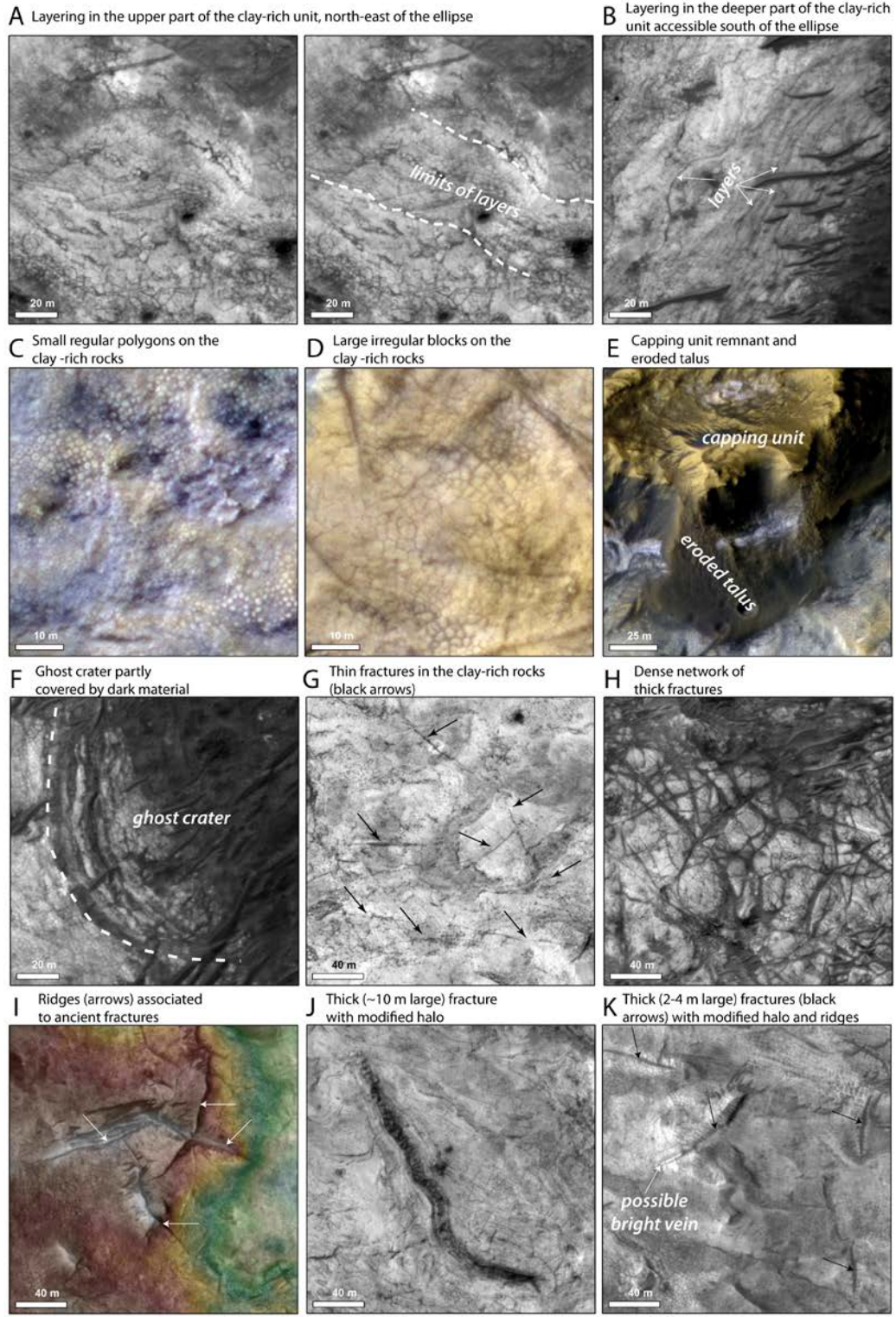

**FIG. 9**

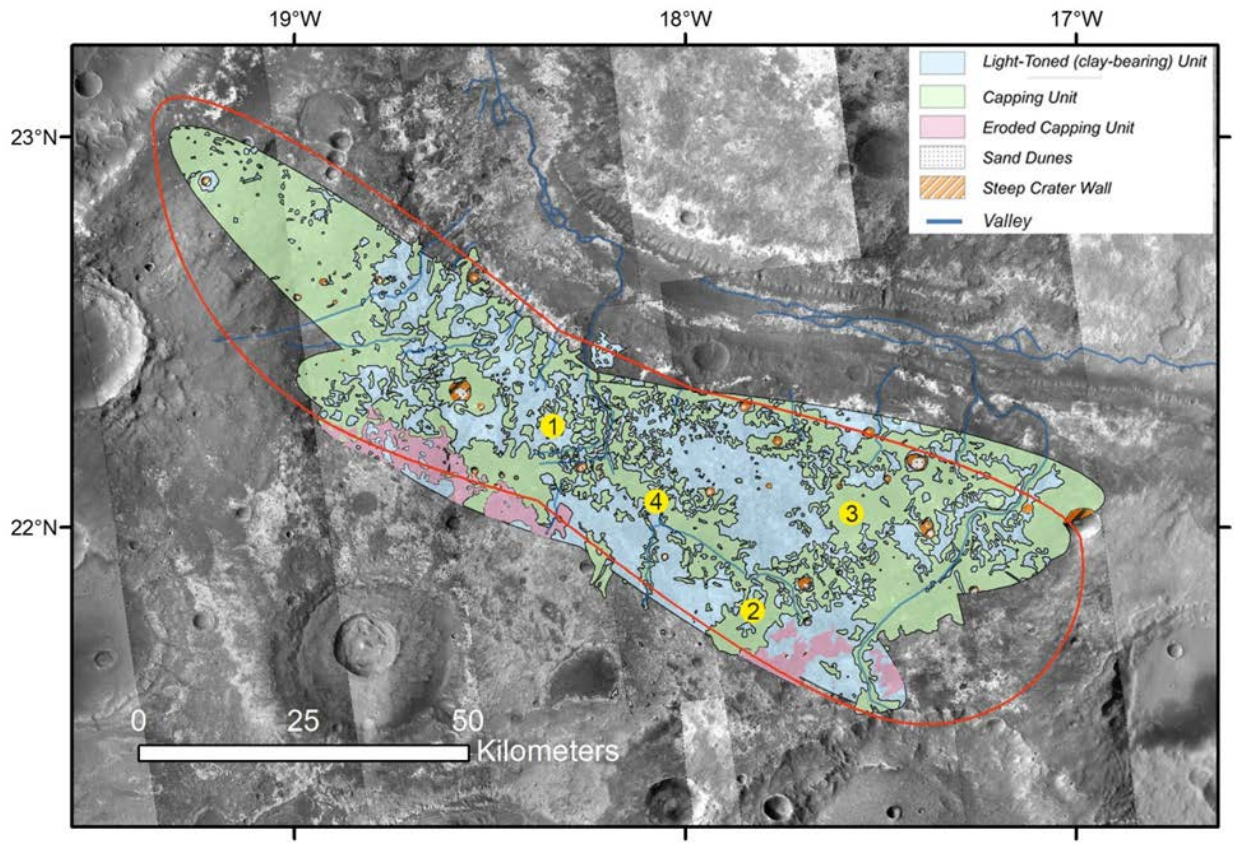

**FIG. 10**

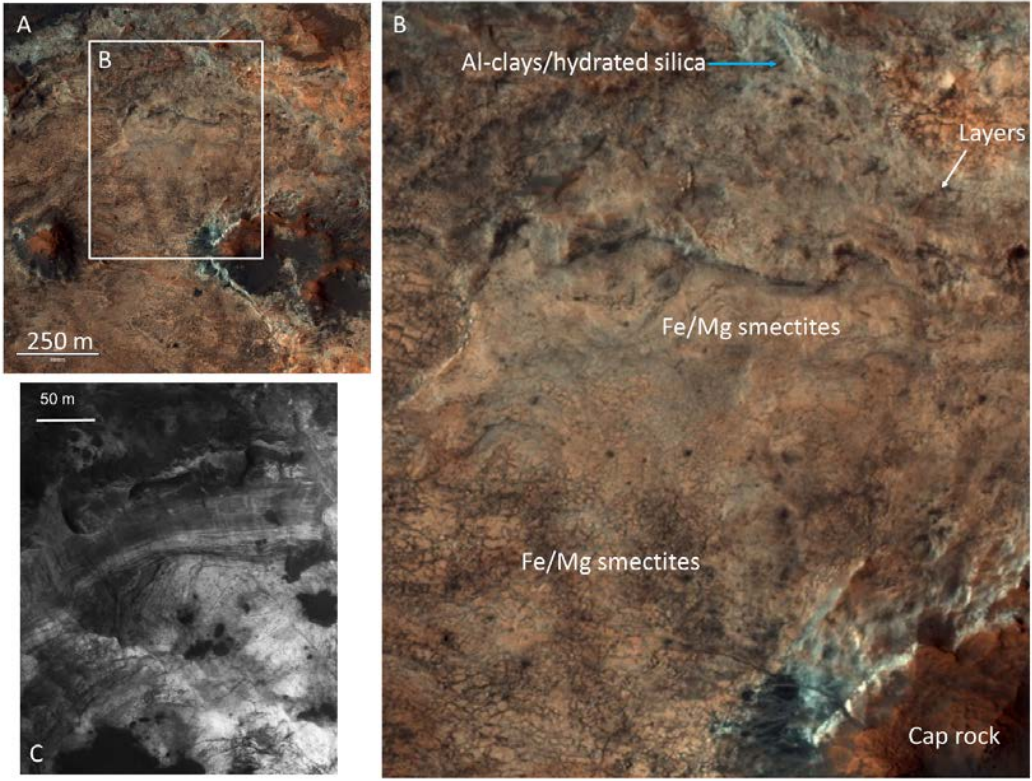

**FIG. 11**

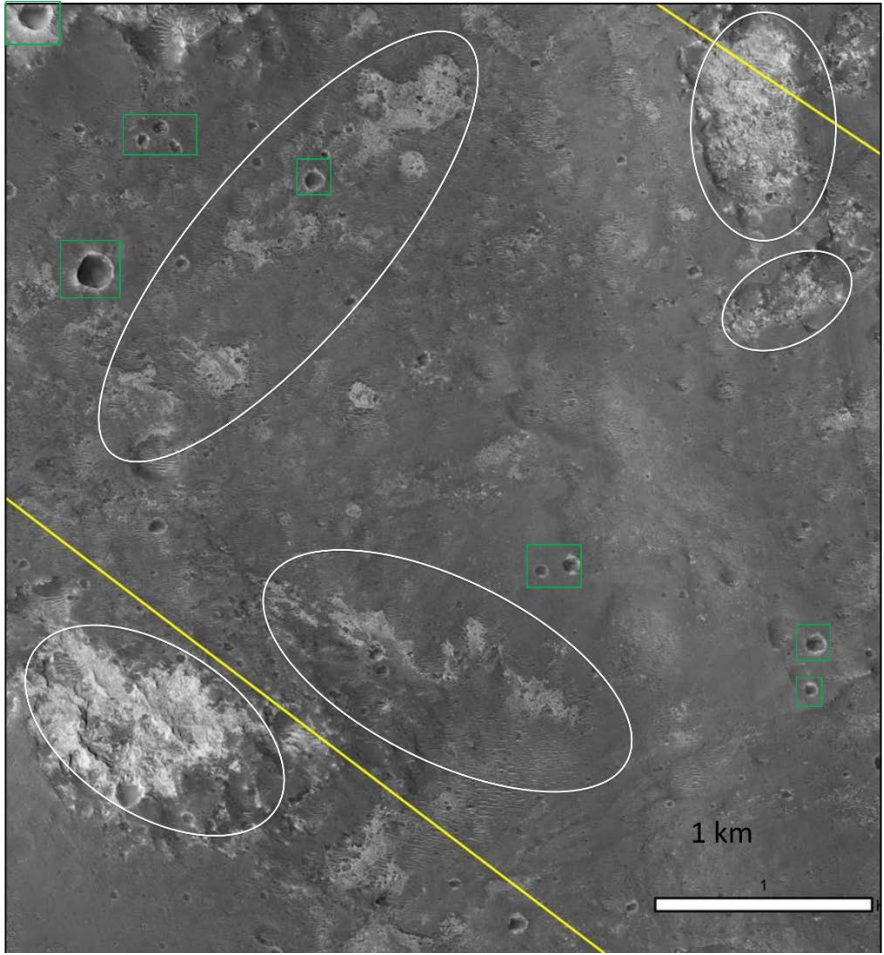

**FIG. 12**

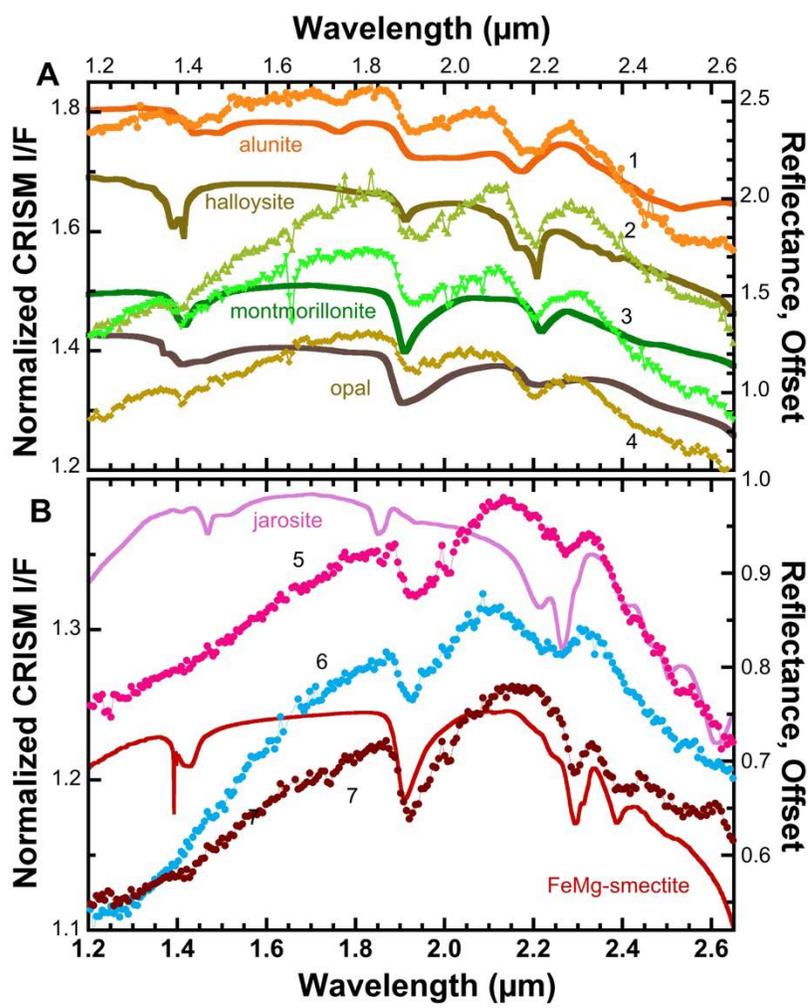

**FIG. 13**

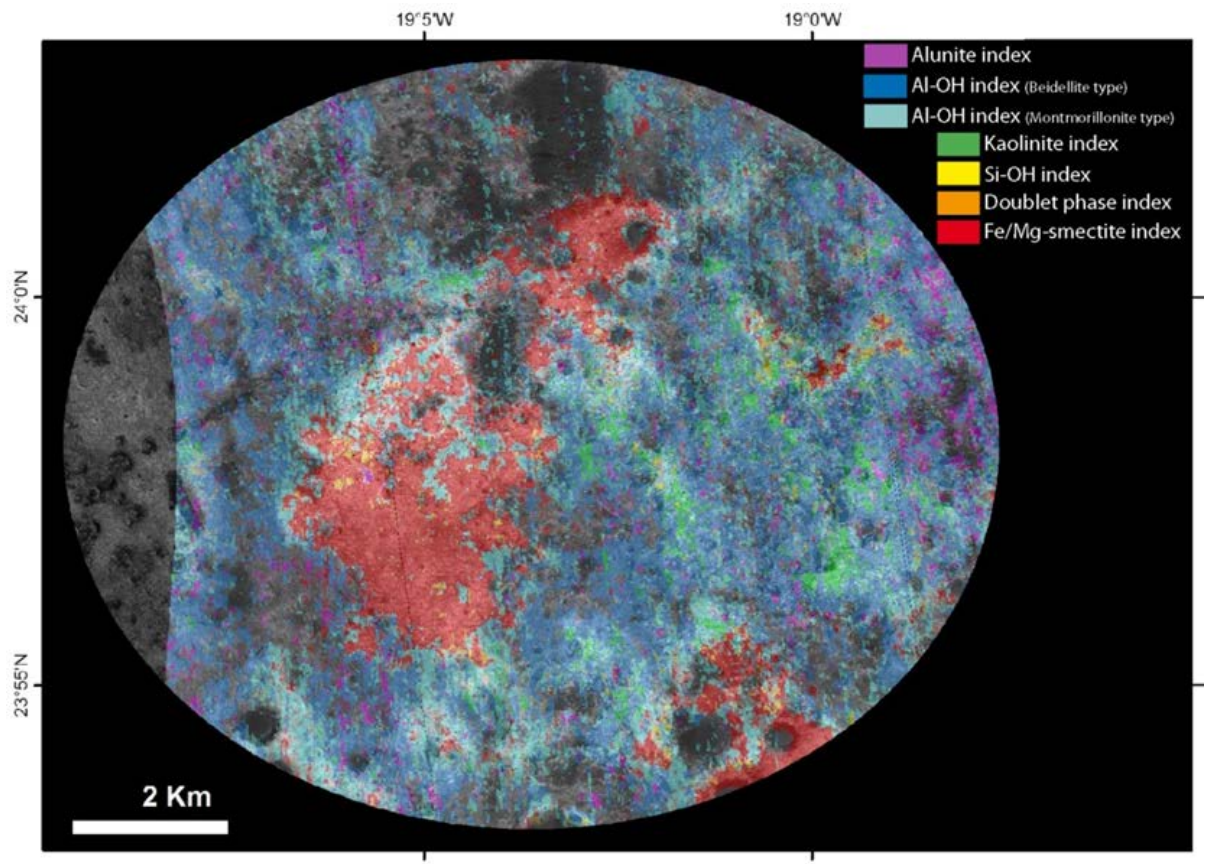

**FIG. 14**

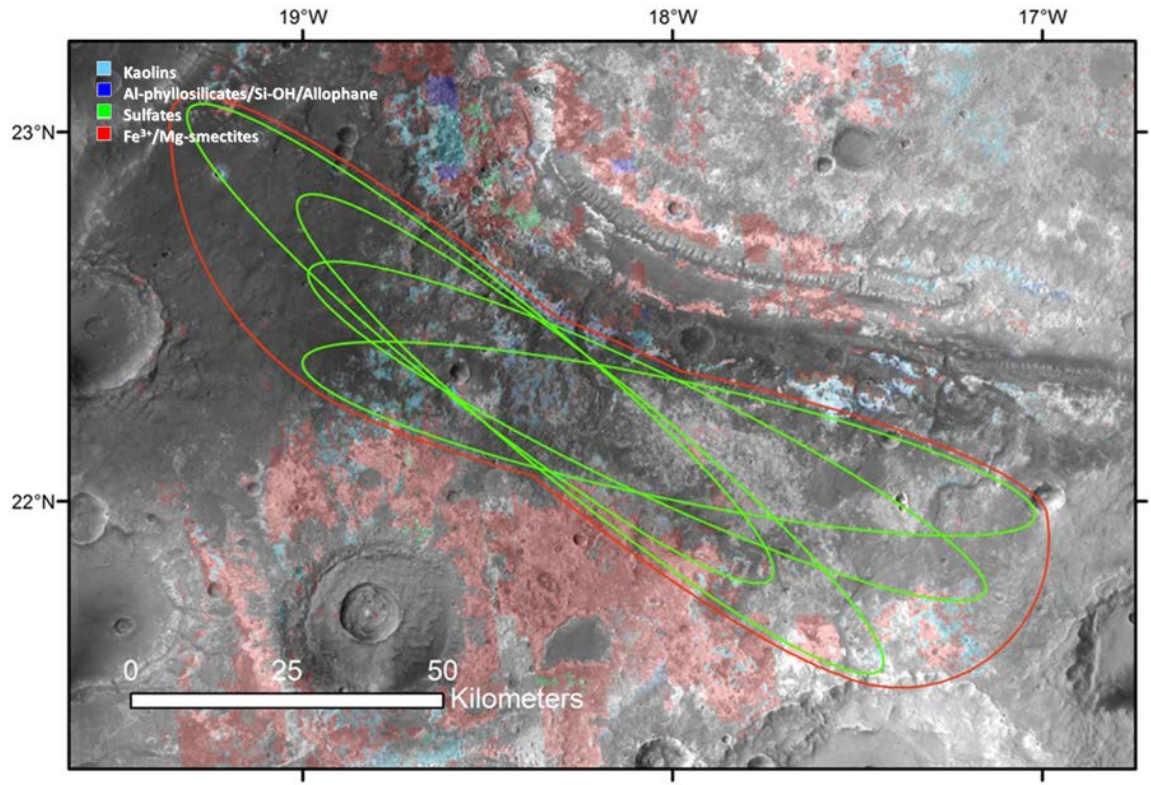

**FIG. 15**

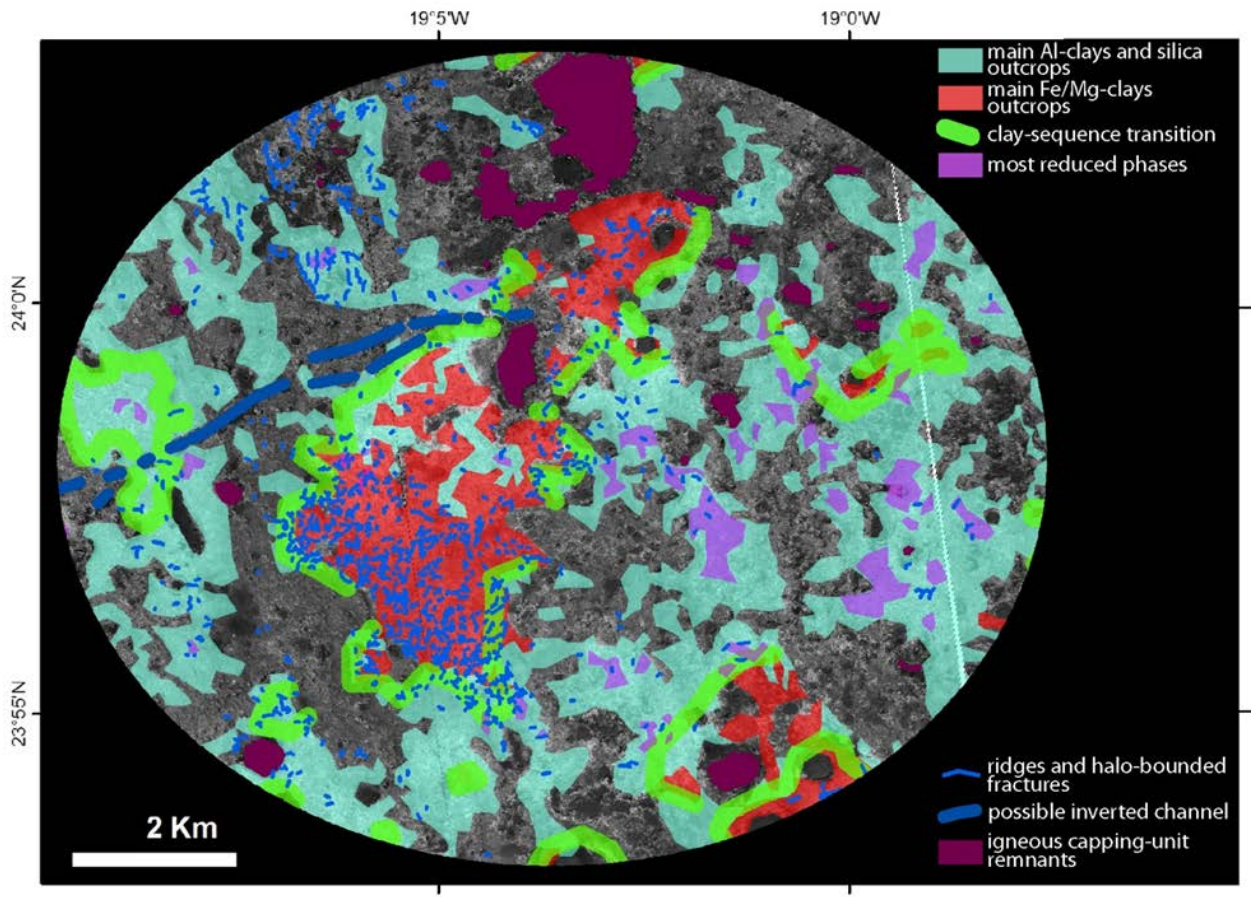

**FIG. 16**

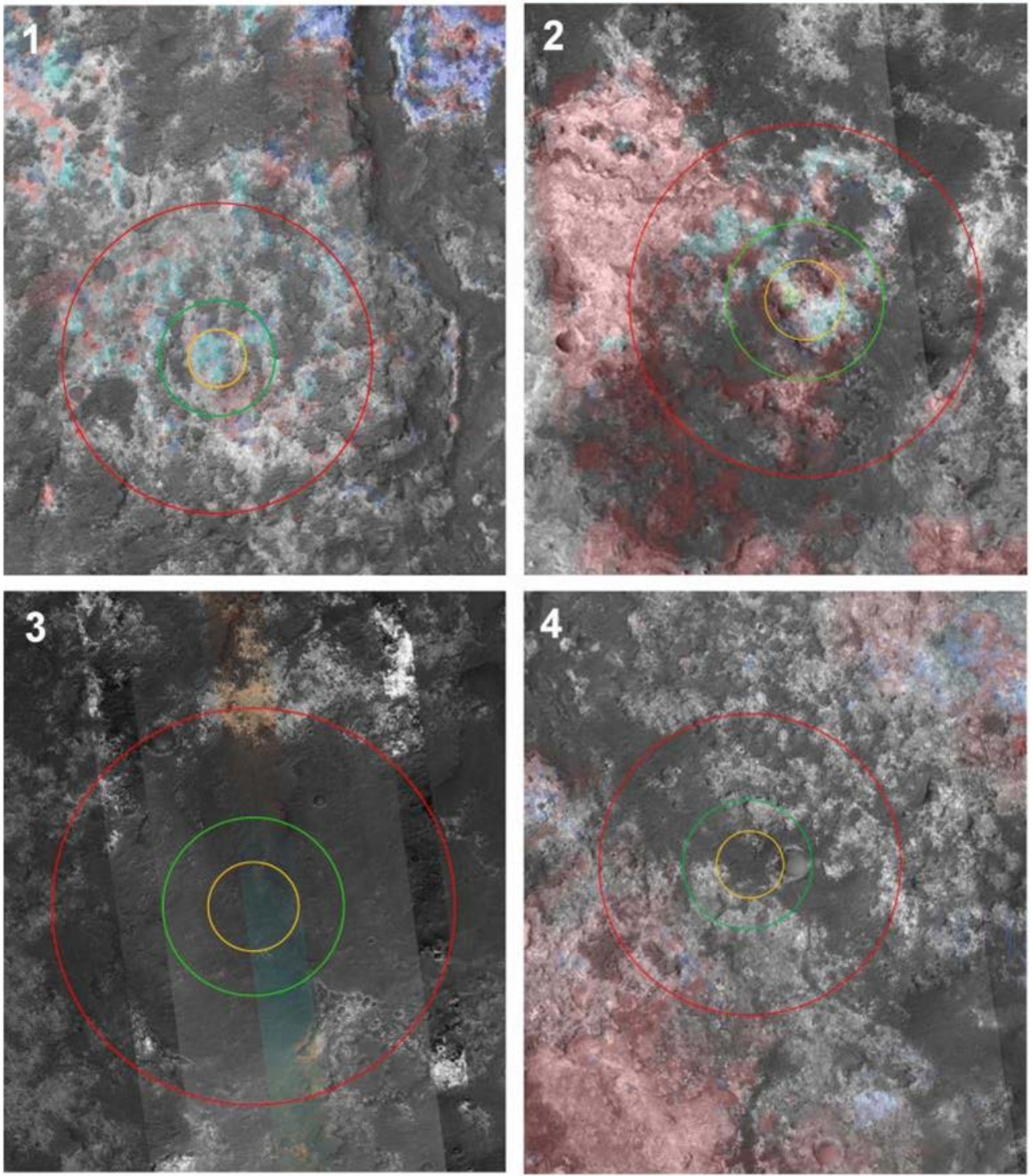

**FIG. 17**

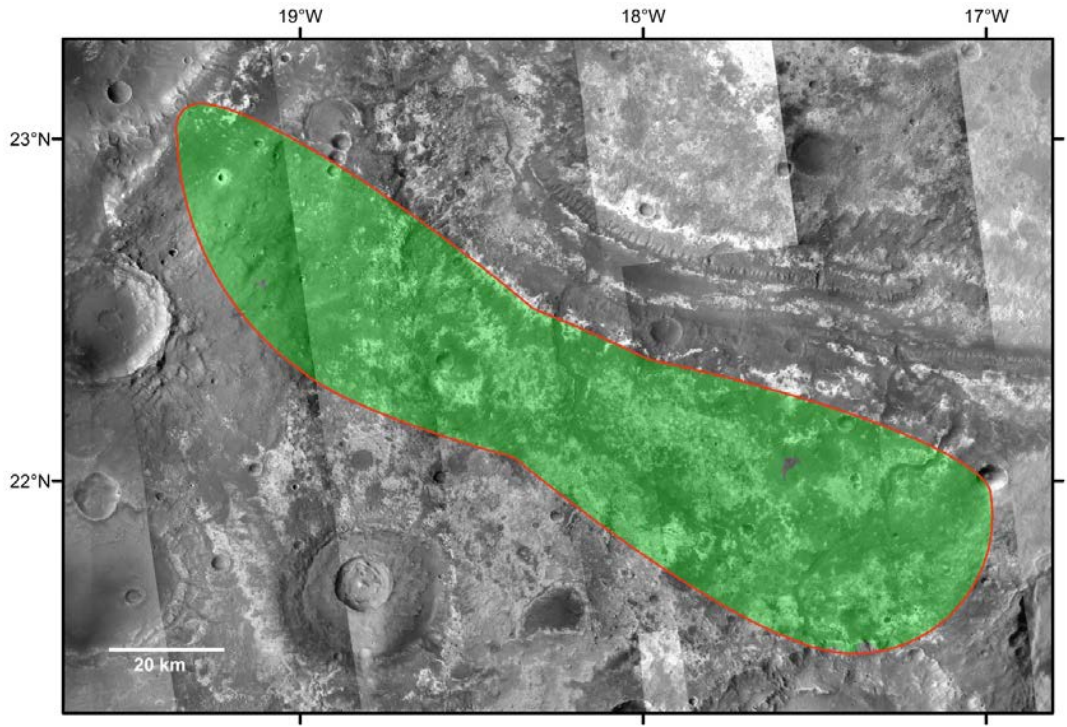

**FIG. 18**

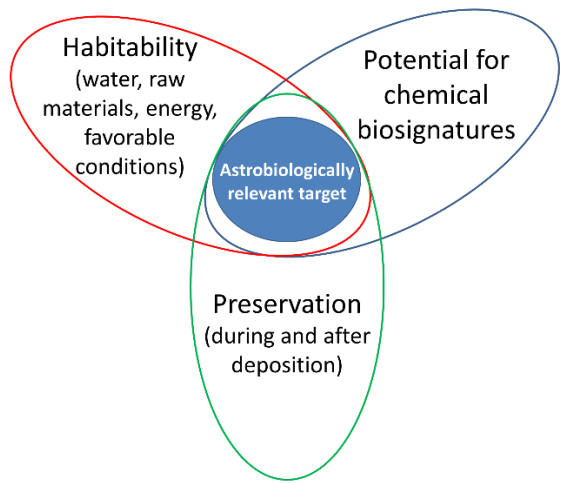

**FIG. 19**

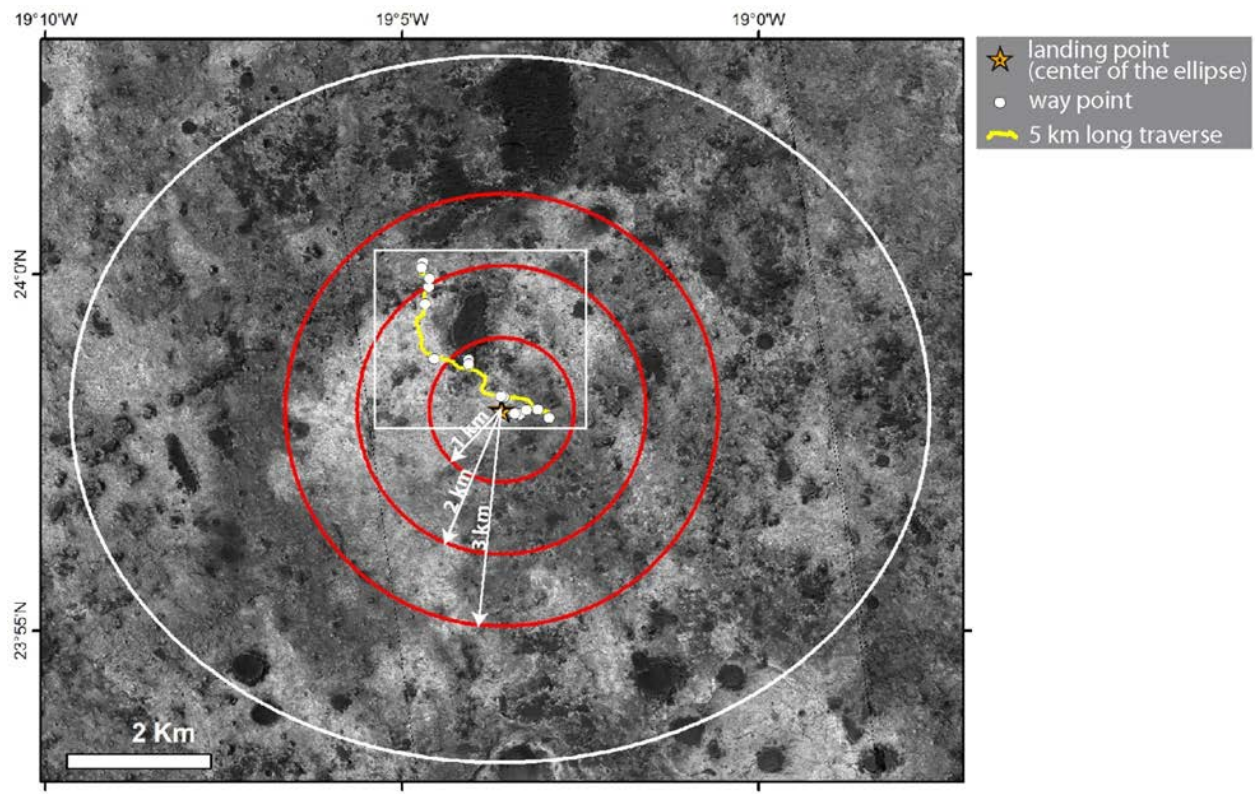

**FIG. 20**

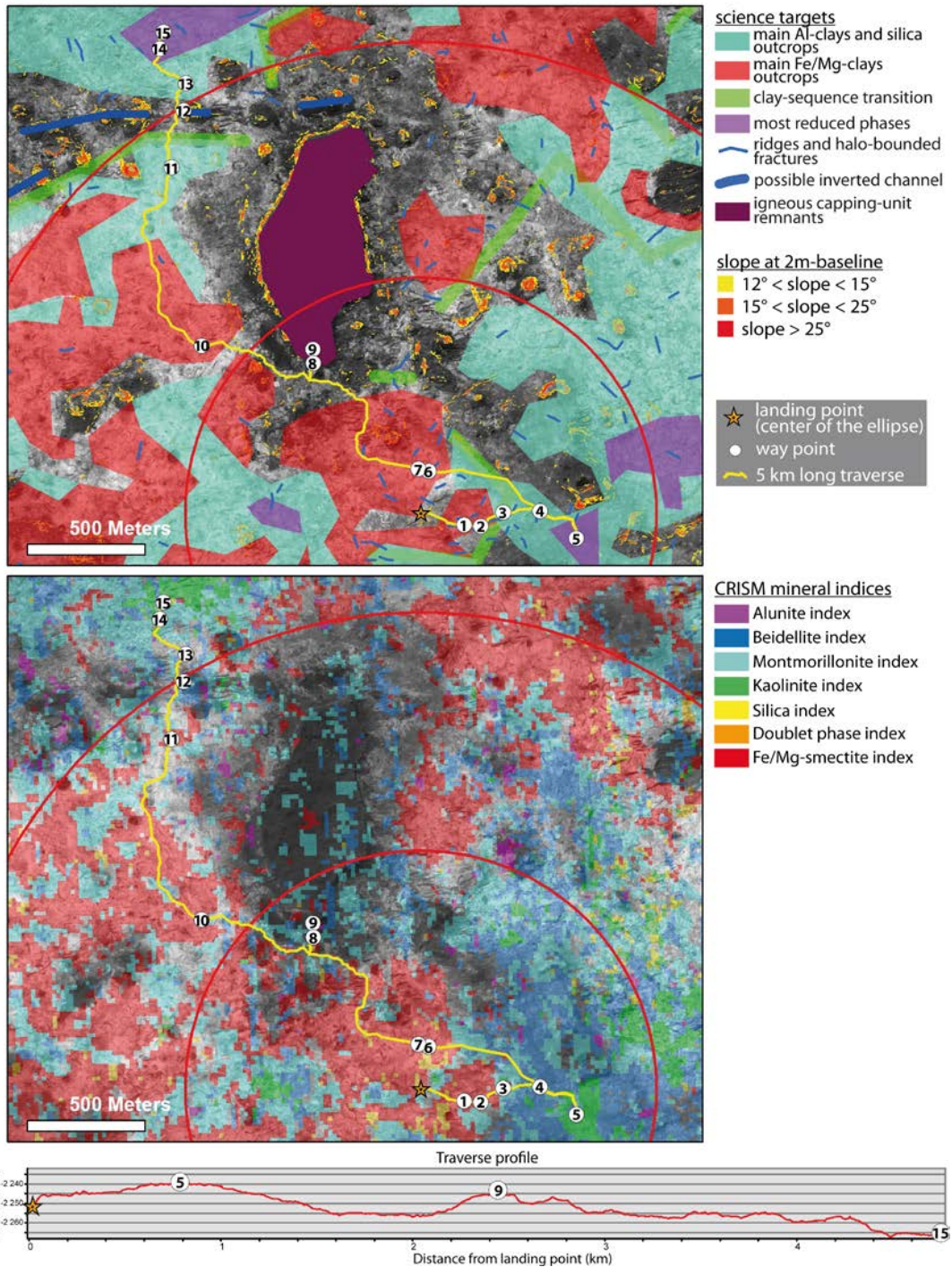

**FIG. 21**

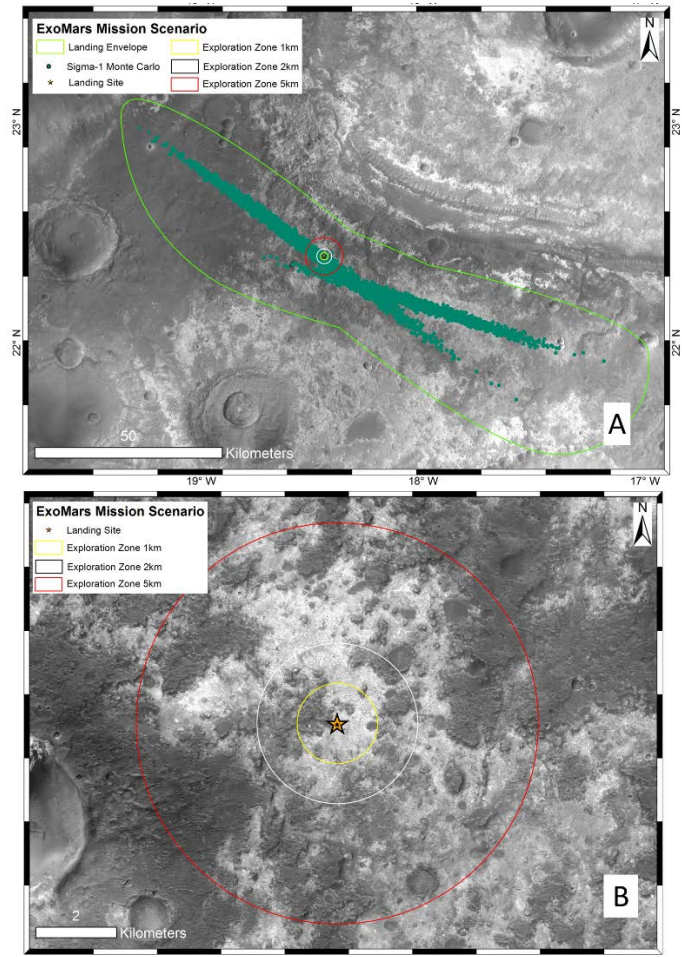

**FIG. 22**

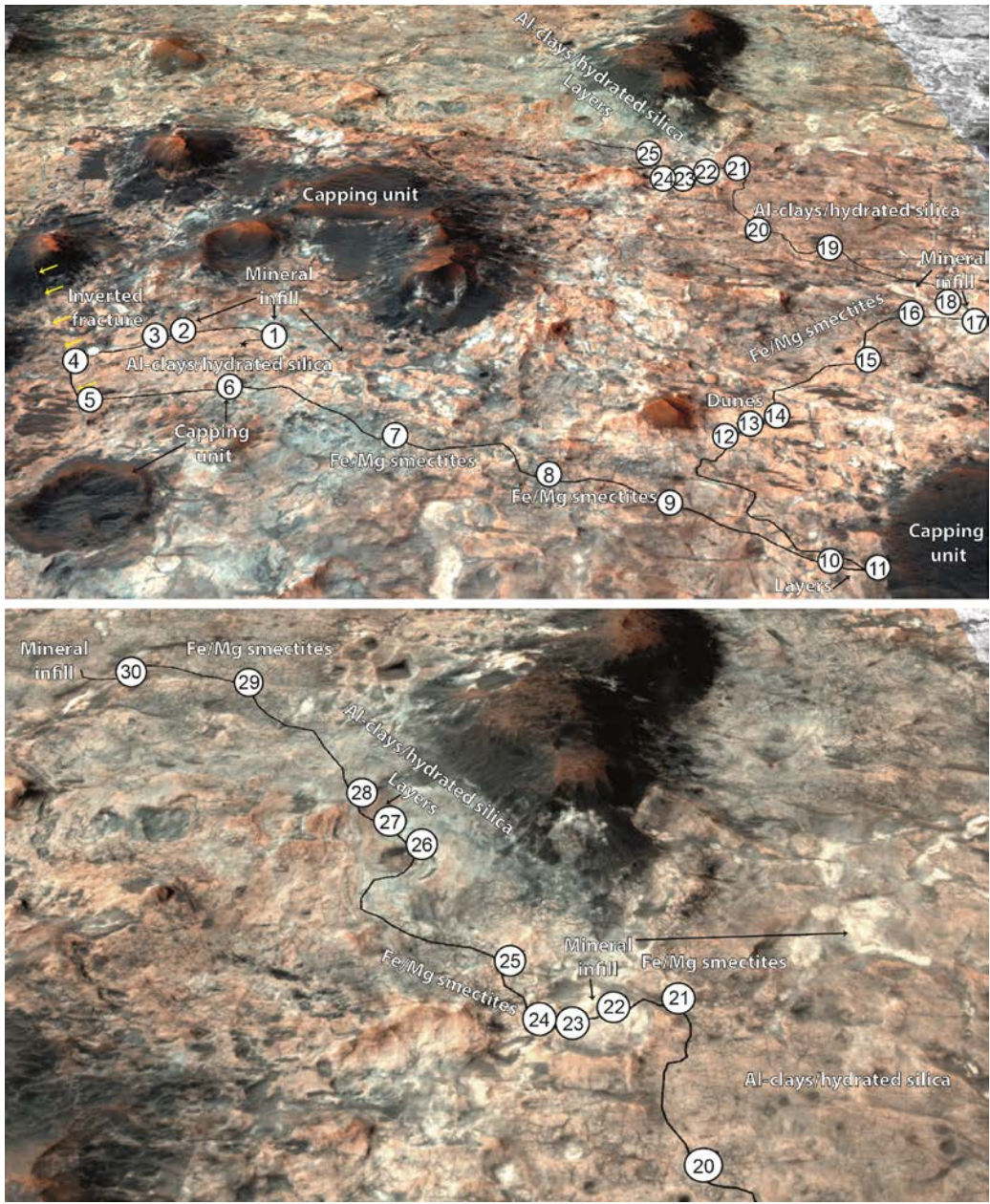

**FIG. 23**

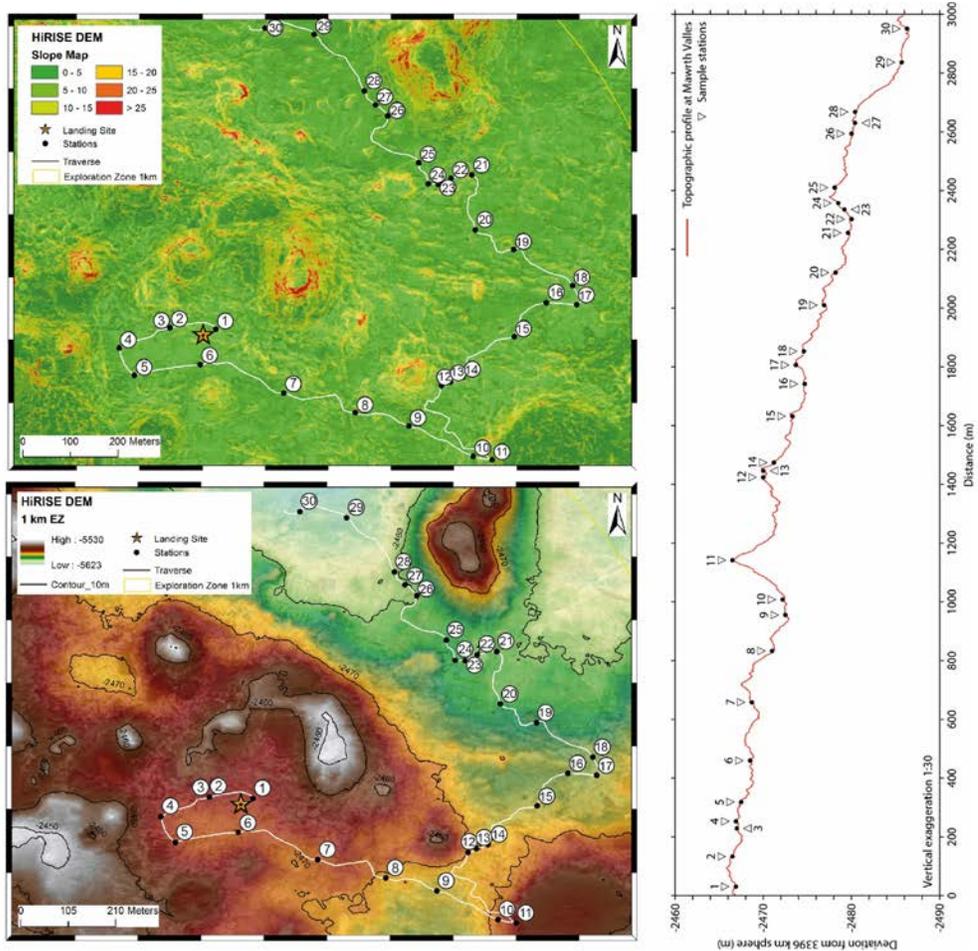

**FIG. 24**

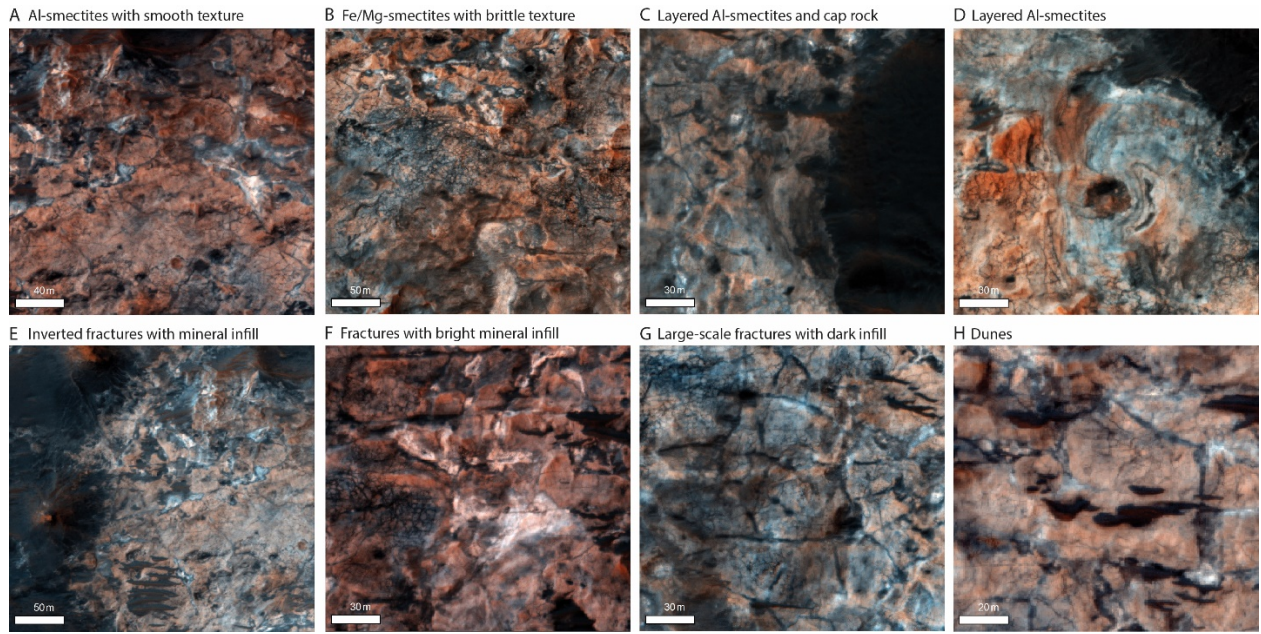

**FIG. 25**